\documentclass[11pt,a4paper]{article}
\pdfoutput=1
\usepackage{jheppub}

\usepackage{placeins}
\usepackage{epsfig}
\usepackage{latexsym}
\usepackage[bbgreekl]{mathbbol}
\usepackage{mathtools}
\usepackage{eqnarray,amsmath,amsfonts,amstext,amssymb,mathrsfs}
\usepackage{multirow}
\usepackage{slashed}
\usepackage{float}
\usepackage{esint}
\usepackage{epsfig}
\usepackage{lscape}
\usepackage{graphicx}
\usepackage{psfrag}
\usepackage[utf8]{inputenc}
\usepackage{subfigure}
\usepackage{nicefrac}
\usepackage[stable]{footmisc}
\usepackage{rotating}
\usepackage{afterpage}
\usepackage{bbm}
\usepackage{color,xcolor}
\usepackage{cancel}

\newcommand{\red}[1]{\textcolor{red}{#1}}
\newcommand{\blue}[1]{\textcolor{blue}{#1}}
\def\){\right)}
\def\({\left( }
\def\]{\right] }
\def\[{\left[ }

\def\NO{\nonumber}

\newcommand{\be}{\begin{equation}}
\newcommand{\ee}{\end{equation}}

\def\bea{\begin{eqnarray}}
\def\eea{\end{eqnarray}}

\def\bal#1\eal{\begin{align}#1\end{align}}

\def\bald{\begin{aligned}}
\def\eald{\end{aligned}}

\def\bsub{\begin{subequations}}
\def\esub{\end{subequations}}

\def\beqx{\begin{displaymath}}
\def\eeqx{\end{displaymath}}

\newcommand{\bmat}{\left(\begin{array}}
\newcommand{\emat}{\end{array}\right)}

\def\a{\alpha}
\def\b{\beta}
\def\c{\chi}
\def\d{\delta}
\def\e{\epsilon}
\def\f{\phi}
\def\g{\gamma}
\def\h{\eta}
\def\j{\psi}
\def\k{\kappa}
\def\l{\lambda}
\def\m{\mu}
\def\n{\nu}
\def\o{\omega}
    
\def\p{\pi}

    \def\th{\theta}
\def\r{\rho}
\def\s{\sigma}
\def\t{\tau}
\def\x{\xi}

\def\F{\Phi}
\def\G{\Gamma}

\def\O{\Omega}

\def\ve{\varepsilon}

\def\ca{{\cal A}}

\def\cd{{\cal D}}
\def\ce{{\cal E}}

\def\cj{{\cal J}}

\def\cn{{\cal N}}
\def\co{{\cal O}}
\def\cp{{\cal P}}
\def\cq{{\cal Q}}

\def\cs{{\cal S}}
\def\ct{{\cal T}}

\def\cw{{\cal W}}

\def\bo{{\raise-.3ex\hbox{\large$\Box$}}}               % D'Alembertian
\def\pa{\partial}                                       % curly d
\def\de{\nabla}                                         % del
\def\face{{\raise.2ex\hbox{$\displaystyle \bigodot$}\mskip-2.2mu \llap {$\ddot
        \smile$}}}                                   % happy face
\def\>{\rangle}                                      %right angle
\def\<{\langle}                                      %left angle

\def\tx#1{\text{#1}}
\def\wt#1{\widetilde{#1}}                            % big tilde
\def\lbar#1{\ensuremath{\overline{#1}}}              % big bar
\def\leftrightarrowfill{$\mathsurround=0pt \mathord\leftarrow \mkern-6mu
        \cleaders\hbox{$\mkern-2mu \mathord- \mkern-2mu$}\hfill
        \mkern-6mu \mathord\rightarrow$}        % <--> double differential
\def\dvec#1{\vbox{\ialign{##\crcr
        \leftrightarrowfill\crcr\noalign{\kern-1pt\nointerlineskip}
        $\hfil\displaystyle{#1}\hfil$\crcr}}}           % <--> accent
\def\diag{{\rm diag \,}}                                % diagonal
\def\-{\hphantom{-}}

\title{Supersymmetry anomalies in $\cn=1$ conformal supergravity}
\author{Ioannis Papadimitriou} 

\affiliation[a]{School of Physics, Korea Institute for Advanced Study, 85 Hoegiro, Seoul 02455, Korea}

\emailAdd{ioannis@kias.re.kr}

\abstract{We solve the Wess-Zumino consistency conditions of $\cn=1$ off-shell conformal supergravity in four dimensions and determine the general form of the superconformal anomalies for arbitrary $a$ and $c$ anomaly coefficients to leading non trivial order in the gravitino. Besides the well known Weyl and $R$-symmetry anomalies, we compute explicitly the fermionic $\cq$- and $\cs$-supersymmetry anomalies. In particular, we show that $\cq$-supersymmetry is anomalous if and only if $R$-symmetry is anomalous. The $\cq$- and $\cs$-supersymmetry anomalies give rise to an anomalous supersymmetry transformation for the supercurrent on curved backgrounds admitting Killing spinors, resulting in a deformed rigid supersymmetry algebra. Our results may have implications for supersymmetric localization and supersymmetry phenomenology. Analogous results are expected to hold in dimensions two and six and for other supergravity theories. The present analysis of the Wess-Zumino consistency conditions reproduces the holographic result of \href{https://arxiv.org/abs/1703.04299}{arXiv:1703.04299} and generalizes it to arbitrary $a$ and $c$ anomaly coefficients.   
}


\preprint{KIAS-P19007}

\begin{document}  
	\maketitle

%\newpage

%\tableofcontents
%\addtocontents{toc}{\protect\setcounter{tocdepth}{3}}
%\renewcommand{\theequation}{\arabic{section}.\arabic{equation}}

\section{Introduction}
\label{intro}
\setcounter{equation}{0}

Supersymmetric quantum field theories have proven invaluable for probing strong coupling physics due to non-renormalization theorems and supersymmetric localization techniques \cite{Witten:1982im,Witten:1988ze,Nekrasov:2002qd,Pestun:2007rz}. They also play a pivotal role in holographic dualities and beyond the Standard Model phenomenology. Given the enormous utility of  supersymmetry, it is rather surprising that the question of whether it is anomalous at the quantum level is still not conclusively answered, despite the extensive literature addressing this question in various contexts. The consensus seems to be that standard supersymmetry (often termed $\cq$-supersymmetry) is not anomalous. However, we demonstrate in this paper that the Wess-Zumino consistency conditions \cite{Wess:1971yu} imply that $\cq$-supersymmetry is necessarily anomalous in theories with an anomalous $R$-symmetry. The same conclusion is reached in the companion paper \cite{Katsianis:2019hhg} by means of an one loop calculation in the free Wess-Zumino model. A related observation was made using the $R$-multiplet in the recent paper \cite{An:2019zok}. These results confirm the $\cq$-supersymmetry anomaly discovered in the context of supersymmetric theories with a holographic dual in \cite{Papadimitriou:2017kzw}, and the related anomalies in rigid supersymmetry \cite{Papadimitriou:2017kzw,An:2017ihs,An:2018roi}.    

Global anomalies do not render the theory inconsistent -- they are a property of the theory and affect physical observables, such as decay channels \cite{Adler:1969gk,Bell:1969ts} and transport coefficients (see \cite{Landsteiner:2016led} for a recent review and \cite{Gooth:2017mbd} for an observation of the mixed chiral-gravitational anomaly in tabletop experiments). They do mean, however, that the theory cannot be coupled consistently to {\em dynamical} gauge fields for the anomalous global symmetry. Specifically, a quantum anomaly in global supersymmetry implies that the theory cannot be coupled consistently to dynamical supergravity at the quantum level. It may also mean that certain conditions necessary to prove non-perturbative results are in fact not met.  

In flat space, global -- or rigid -- anomalies are typically visible only in higher-point functions as contact terms that violate the classical Ward identities. For example, the lowest correlation functions where the $\cq$-supersymmetry anomaly is visible in flat space are four-point functions involving two supercurrents and either two $R$-currents or one $R$-current and one stress tensor \cite{Katsianis:2019hhg}. However, global anomalies become manifest at the level of the quantum effective action and in one-point functions when arbitrary sources for the current operators are turned on (i.e. when the theory is coupled to {\em background} gauge fields for the global symmetries), or when the theory is put on a curved background admitting Killing symmetries. In particular, global supersymmetry anomalies are related to supersymmetric index theorems and may affect observables such as partition functions, the Casimir energy, and Wilson loop expectation values of supersymmetric theories on curved backgrounds admitting rigid supersymmetry. 

Following the recent advances in supersymmetric localization techniques \cite{Pestun:2007rz} (see \cite{Pestun:2016zxk} for a comprehensive review), supersymmetric quantum field theories on curved backgrounds have attracted considerable interest. A systematic procedure for placing a supersymmetric theory on a curved background was proposed in \cite{Festuccia:2011ws}. The first step is coupling the theory to a given off-shell background supergravity, which corresponds to turning on arbitrary sources for the current multiplet operators and promoting the global symmetries -- including supersymmetry -- to local ones. The Killing spinor equations obtained by setting the supersymmetry variations of the fermionic background fields to zero determine the curved backgrounds that admit a notion of rigid supersymmetry. Such backgrounds have been largely classified for a number of off-shell supergravity theories and for various spacetime dimensions \cite{Samtleben:2012gy,Klare:2012gn,Dumitrescu:2012ha,Liu:2012bi,Dumitrescu:2012at,Kehagias:2012fh,Closset:2012ru,Samtleben:2012ua,Cassani:2012ri,deMedeiros:2012sb,Hristov:2013spa} (see also \cite{Blau:2000xg,Kuzenko:2012vd} for earlier work). However, this procedure is classical and does not account for possible quantum anomalies. 

It was in that context that the anomalies in  $\cq$-supersymmetry \cite{Papadimitriou:2017kzw} and rigid supersymmetry \cite{Papadimitriou:2017kzw,An:2017ihs} were discovered, providing a resolution to an apparent tension between the field theory analysis of \cite{Closset:2013vra,Closset:2014uda,Assel:2014paa} and the holographic result of \cite{Genolini:2016ecx}. Based on the classical supersymmetry algebra on curved backgrounds that admit a certain number of supercharges, the authors of \cite{Closset:2013vra,Closset:2014uda,Assel:2014paa} demonstrated that the supersymmetric partition function on such backgrounds should be independent of specific deformations of the supersymmetric background. However, an explicit evaluation of the on-shell action of minimal $\cn=2$ gauged supergravity on supersymmetric asymptotically locally AdS$_5$ solutions using holographic renormalization \cite{Henningson:1998gx,deHaro:2000vlm} in \cite{Genolini:2016ecx} demonstrated that the holographic partition function on the same supersymmetric backgrounds does in fact depend on the deformation parameters. The resolution to this apparent contradiction was provided in \cite{Papadimitriou:2017kzw}, where both the bosonic and fermionic superconformal Ward identities were derived holographically, including the corresponding superconformal anomalies. It was then shown that the anomalies in the fermionic Ward identities (the ones in the divergence and the gamma-trace of the supercurrent) lead to a deformed superconformal algebra on backgrounds admitting Killing spinors. Repeating the argument of \cite{Closset:2013vra,Closset:2014uda,Assel:2014paa} using this deformed supersymmetry algebra reproduced exactly the dependence on the deformations of the supersymmetric background seen in \cite{Genolini:2016ecx}.\footnote{At least for theories with $a=c$ on supersymmetric backgrounds of the form $S^1\times M_3$ with $M_3$ a Seifert manifold it is possible to remove the rigid supersymmetry anomaly at the expense of breaking certain diffeomorphisms using the local but non-covariant counterterm found in \cite{Genolini:2016ecx}. We will elaborate on this point in section \ref{supercurrent-anomalies}.}      

The Ward identities and quantum anomalies of four dimensional superconformal theories have been studied extensively over the years \cite{McArthur:1983fk,Bonora:1984pn,Buchbinder:1986im,Brandt:1993vd,Anselmi:1997am,Piguet:1998bj,Erdmenger:1998ew,Erdmenger:1998xv,Bonora:2013rta,Butter:2013ura,Cassani:2013dba,Auzzi:2015yia}. They are usually discussed in superspace language and can be written compactly in the form (see e.g. appendix A of \cite{Anselmi:1997am})
\be\label{superspaceWI}
\nabla^{\dot\a}J_{\a\dot\a}=\nabla_\a J,
\ee  
where the supertrace superfield $J$ is given by
\be\label{superspace-anomaly-0}
J=\frac{1}{24\p^2}\big(c\cw^2-a\ce\big),
\ee
and 
\be
\cw^2=\frac12 W_{\a\b\g}W^{\a\b\g}, \qquad \ce=\cw^2+(\lbar\nabla^2+R)(G^2+2\lbar R R),
\ee
are respectively the square of the superWeyl tensor and the chirally projected superEuler density. The chiral superfields $W_{\a\b\g}$ and $R$ and the vector superfield $G_{\a\dot\a}$ are the three superspace curvatures \cite{Wess:1992cp} and in the conventions of \cite{Anselmi:1997am} $G^2=\frac12 G^{\a\dot\a}G_{\a\dot\a}$. The components of the supertrace superfield $J$ contain the trace of the stress tensor, the gamma-trace of the supercurrent, and the divergence of the $R$-current. The divergence of the stress tensor and of the supercurrent appear as components of the superspace conservation equation \eqref{superspaceWI}. 

The superspace analyses of \cite{Bonora:1984pn, Bonora:2013rta} and \cite{Brandt:1993vd} seem especially related to our results in this paper. In particular, the anomalies in the divergence and in the gamma-trace of the supercurrent we derive are likely related to the fermionic components of the superspace cocycles found in \cite{Bonora:1984pn,Bonora:2013rta,Brandt:1993vd}, even though none of these earlier works concerns $\cn=1$ conformal supergravity and so the field content is somewhat different. Moreover, to the best of our knowledge these fermionic components have not been written explicitly in the literature before and so a direct comparison with our results is not  straightforward. Another result that may be related to the anomalies we find here is \cite{Auzzi:2015yia}, where it was shown that in the presence of anomalous Abelian flavor symmetries the Wess-Zumino consistency conditions require additional -- non holomorphic -- terms to the supertrace anomaly. These terms seem related to terms we find here in the case of an anomalous $R$-symmetry. 

Finally, we should mention that there is an extensive body of literature discussing supersymmetry anomalies in the presence of gauge anomalies in supersymmetric gauge theories, reviewed in \cite{Piguet:1986ug}. Such anomalies involve the dynamical fields in the Lagrangian description of the gauge theory in flat space and are distinct from the supersymmetry anomalies we identify in the present paper, which involve the background supergravity fields. However, the mathematical structure underlying the descent equations that relates the $R$-symmetry and supersymmetry anomalies is identical to that relating gauge anomalies to supersymmetry anomalies \cite{Itoyama:1985qi} (see also \cite{Guadagnini:1985ea,Piguet:1984aa}). 

In this paper we consider $\cn=1$ off-shell conformal supergravity in four dimensions 
\cite{Kaku:1977pa,Kaku:1977rk,Kaku:1978nz,Townsend:1979ki}, which provides a suitable background for superconformal theories via the construction of \cite{Festuccia:2011ws}. We determine the algebra of local symmetry transformations  and derive the corresponding classical Ward identities. The main result of the paper is the solution of the Wess-Zumino consistency conditions \cite{Wess:1971yu} associated with the $\cn=1$ conformal supergravity algebra from which we obtain the general form of the superconformal anomalies to leading non trivial order in the gravitino for arbitrary $a$ and $c$ anomaly coefficients. Our analysis is carried out in components and we explicitly determine the fermionic Ward identities corresponding to the divergence and the gamma-trace of the supercurrent, including their anomalies. To the best of our knowledge these have not appeared in the literature before, at least explicitly. The bosonic Ward identities and anomalies reproduce well known results \cite{Anselmi:1997am} (corrected in \cite{Cassani:2013dba}). We find that the divergence of the supercurrent, which corresponds to the Ward identity associated with $\cq$-supersymmetry, is anomalous whenever $R$-symmetry is anomalous. Moreover, the $\cn=1$ supergravity algebra dictates that the $\cq$-supersymmetry anomaly cannot be removed by a local counterterm without breaking diffeomorphisms and/or local Lorentz transformations. The Ward identities and the superconformal anomalies we obtain by solving the Wess-Zumino conditions reproduce those found holographically in \cite{Papadimitriou:2017kzw} in the special case when the $a$ and $c$ anomaly coefficients are equal.  

The paper is organized as follows. In section \ref{algebra} we review relevant aspects of $\cn=1$ off-shell conformal supergravity and determine the algebra of its local symmetry transformations. These transformations are used in section \ref{CWIDs} to derive the corresponding classical Ward identities. The main result is presented in section \ref{WZanomalies}, where we obtain the general form of the superconformal anomalies by solving the Wess-Zumino consistency conditions associated with the $\cn=1$ conformal supergravity algebra. The actual calculation is shown in considerable detail in appendix \ref{WZ-proof}. In section \ref{supercurrent-anomalies} we determine the anomalous transformation of the supercurrent under local supersymmetry and discuss the implications for the rigid supersymmetry algebra on curved backgrounds admitting Killing spinors of conformal supergravity. We conclude in section \ref{conclusion} and collect our conventions and several gamma matrix identities in appendix \ref{conventions}.

\section{The local symmetry algebra of $\cn=1$ conformal supergravity}
\label{algebra}
\setcounter{equation}{0}

In this section we review relevant aspects of $\cn=1$ off-shell conformal supergravity in four dimensions and determine its local off-shell symmetry algebra as a preparatory step for solving the Wess-Zumino consistency conditions. $\cn=1$ conformal supergravity can be constructed as a gauge theory of the superconformal algebra \cite{Kaku:1977pa,Kaku:1977rk,Kaku:1978nz,Townsend:1979ki} (see \cite{VanNieuwenhuizen:1981ae,deWit:1981vgr,deWit:1983qkc,Fradkin:1985am} and chapter 16 of \cite{Freedman:2012zz} for pedagogical reviews). Its field content consists of the vielbein $e^a_\m$, an Abelian gauge field $A_\m$, and a Majorana gravitino $\j_{\m}$, which comprise 5+3 bosonic and 8 fermionic off-shell degrees of freedom. 

The reason for focusing on $\cn=1$ conformal supergravity here is threefold. Firstly, background conformal supergravity is relevant for describing the Ward identities of superconformal theories and their quantum anomalies. Moreover, other supergravity theories can be obtained from conformal supergravity by coupling it to compensator  multiplets and gauge fixing via the so called tensor or multiplet calculus \cite{Kaku:1978ea,Stelle:1978yr,deWit:1983qkc}. When applied to background supergravity, this procedure may be thought of as the process of turning on local relevant couplings at the ultraviolet superconformal fixed point. Finally,       off-shell $\cn=1$ conformal supergravity in four dimensions is induced on the conformal boundary of five dimensional anti de Sitter space by minimal $\cn=2$ on-shell gauged supergravity in the bulk \cite{Balasubramanian:2000pq}. This means that the Wess-Zumino consistency conditions for $\cn=1$ conformal supergravity should reproduce the superconformal anomalies obtained holographically in \cite{Papadimitriou:2017kzw} for the case $a=c$. We will see in the subsequent sections that this is indeed the case.   

In the construction of $\cn=1$ conformal supergravity as a gauge theory of the superconformal algebra, $\cq$- and $\cs$-supersymmetry are on the same footing before the curvature constraints are imposed, each having its own independent gauge field, respectively $\j_\m$ and $\f_\m$. The covariant derivative acts on these gauge fields as
\bal\label{covD-psi-phi}
\cd_\m\j_\n\equiv&\;\Big(\pa_\m+\frac14\o_\m{}^{ab}(e,\j)\g_{ab}+i\g^5A_\m\Big)\j_\n-\G^\r_{\m\n}\j_\r\equiv \big(\mathscr{D}_\m+i\g^5A_\m\big)\j_\n,\NO\\
\cd_\m\f_\n=&\;\Big(\pa_\m+\frac14\o_\m{}^{ab}(e,\j)\g_{ab}-i\g^5A_\m\Big)\f_\n-\G^\r_{\m\n}\f_\r=\big(\mathscr{D}_\m-i\g^5A_\m\big)\f_\n,
\eal
where $\o_\m{}^{ab}(e,\j)$ denotes the torsion-full spin connection\footnote{The purely bosonic part of the spin connection, $\o_\m{}^{ab}(e)$, is torsion-free. For most part of the subsequent analysis we will work to leading non trivial order in the gravitino, in which case the torsion-free part of the spin connection suffices. However, the full spin connection is necessary in order to determine e.g. the local symmetry algebra.} 
\be
\o_\m{}^{ab}(e,\j)\equiv\o_\m{}^{ab}(e)+\frac14\big(\lbar\j_a\g_\m\j_b+\lbar\j_\m\g_a\j_b-\lbar\j_\m\g_b\j_a\big).
\ee
Once the curvature constraints are imposed, however, the gauge field $\f_\m$ ceases to be an independent field and it is expressed locally in terms of the physical fields as
\be\label{phi}
\f_\m\equiv \frac13\g^\n\Big(\cd_\n\j_\m-\cd_\m\j_\n-\frac{i}{2}\g^5 \e_{\n\m}{}^{\r\s}\cd_\r\j_\s\Big)=-\frac16\big(4\d^{[\r}_\m\d^{\s]}_\n+i\g^5 \e_{\m\n}{}^{\r\s}\big)\g^\n \cd_\r\j_\s.
\ee
As we will see shortly, this quantity appears in the supersymmetry transformation of the gauge field $A_\m$ as well as in the fermionic superconformal anomalies. 

Our spinor conventions are given in appendix \ref{conventions} and follow those of \cite{Freedman:2012zz}. Compared to \cite{Fradkin:1985am}, we use Lorentzian signature instead of Euclidean and we have rescaled the gauge field $A_\m$ according to $-\frac34 A^{\rm FT}_\m\to A_\m$ in order for its coefficient in the covariant derivatives \eqref{covD-psi-phi} to be unity, as is standard in the field theory literature.

\subsubsection*{Local symmetry transformations}

The local symmetries of $\cn=1$ conformal supergravity are diffeomorphisms $\x^\m(x)$, Weyl transformations $\s(x)$, local frame rotations $\l^{ab}(x)$, $U(1)$ gauge transformations $\th(x)$, as well as $\cq$- and $\cs$-supersymmetry, parameterized respectively by the local spinors $\ve(x)$ and $\h(x)$. The covariant derivative acts on the spinor parameters $\ve$ and $\h$ as
\bal\label{covD-parameters}
\cd_\m\ve\equiv&\;\Big(\pa_\m+\frac14\o_\m{}^{ab}(e,\j)\g_{ab}+i\g^5A_\m\Big)\ve\equiv \big(\mathscr{D}_\m+i\g^5A_\m\big)\ve,\NO\\
\cd_\m\h\equiv&\;\Big(\pa_\m+\frac14\o_\m{}^{ab}(e,\j)\g_{ab}-i\g^5A_\m\Big)\h\equiv \big(\mathscr{D}_\m-i\g^5A_\m\big)\h.
\eal

Under these local transformations the fields of $\cn=1$ conformal supergravity transform as   
\bal\label{sugra-trans}
\d e^a_\m=&\;\x^\l\pa_\l e^a_\m+e^a_\l\pa_\m\x^\l-\l^a{}_b e^b_\m+\s e^a_\m-\frac12\lbar\j_\m\g^a\ve,\NO\\
\d\j_\m=&\;\x^\l\pa_\l\j_\m+\j_\l\pa_\m\x^\l-\frac14\l_{ab}\g^{ab}\j_\m+\frac12\s\j_\m+\cd_\m\ve-\g_\m\h- i\g^5\th\j_\m,\NO\\
\d A_\m=&\;\x^\l\pa_\l A_\m+A_\l\pa_\m\x^\l+\frac{3i}{4}\lbar\f_\m\g^5\ve-\frac{3i}{4}\lbar\j_\m\g^5\h+\pa_\m\th.
\eal
These transformations imply that the quantity $\f_\m$ in \eqref{phi} transforms as
\be\label{phi-trans}
\d\f_\m=\x^\l\pa_\l\f_\m+\f_\l\pa_\m\x^\l-\frac14\l_{ab}\g^{ab}\f_\m-\frac12\s\f_\m+\frac12\Big(P_{\m\n}+\frac{2i}{3}F_{\m\n}\g^5-\frac{1}{3}\wt F_{\m\n}\Big)\g^\n\ve+\cd_\m\h+i\g^5\th\f_\m,
\ee
where
\be\label{Schouten}
P_{\m\n}\equiv\frac12\Big(R_{\m\n}-\frac16Rg_{\m\n}\Big),
\ee
is the Schouten tensor in four dimensions and the dual fieldstrength is defined as 
\be\label{dualF}
\wt F_{\m\n}\equiv\frac12 \e_{\m\n}{}^{\r\s}F_{\r\s}.
\ee

Notice that the transformations \eqref{sugra-trans} coincide with those induced on the boundary of five dimensional anti de Sitter space by minimal $\cn=2$ gauged supergravity in the bulk \cite{Balasubramanian:2000pq,Papadimitriou:2017kzw}. In order to compare with the results of \cite{Papadimitriou:2017kzw} one should take into account that we have rescaled the gauge field and the local symmetry parameters according to  
\be
\sqrt{3}A_\m^{\text{there}}/\ell\to A_\m,
\ee
and
\be
\s_{\text{there}}/\ell\to\s,\qquad \e_+^{\text{there}}/\ell\to\ve,\qquad \e_-^{\text{there}}/\ell\to\h,\qquad \sqrt{3}\;\th_{\text{there}}/\ell\to\th,
\ee
where ``there'' refers to the variables used in \cite{Papadimitriou:2017kzw}. Moreover, we use the Majorana formulation of $\cn=1$ conformal supergravity here instead of the Weyl formulation used in \cite{Papadimitriou:2017kzw}.

\subsubsection*{Local symmetry algebra}

The local transformations \eqref{sugra-trans} determine the algebra of local symmetries, i.e. the commutators $[\d_\O,\d_{\O'}]$, where $\O$ and $\O'$ denote any of the local parameters $\s,\x,\l,\th,\ve,\h$. Off-shell closure of the algebra requires that the parameters transform under the local symmetries as
\bal\label{param-trans}
&\d\x^\m=\x'^\n\pa_\n\x^\m-\x^\n\pa_\n\x'^\m,\qquad 
\d\l^a{}_b=\x^\m\pa_\m\l^a{}_b,\qquad
\d\s=\x^\m\pa_\m\s,\qquad
\d\th=\x^\m\pa_\m\th,\NO\\
&\d\ve=\x^\m\pa_\m\ve+\frac12\s\ve-\frac14\l_{ab}\g^{ab}\ve-i\th\g^5\ve,\qquad 
\d\h=\x^\m\pa_\m\h-\frac12\s\h-\frac14\l_{ab}\g^{ab}\h+i\th\g^5\h.
\eal
Applying the transformations \eqref{sugra-trans} repeatedly we then find that (to leading order in the gravitino) the only non vanishing commutators and the corresponding composite symmetry parameters are:
\begin{alignat}{3}\label{CS-local-algebra}
&\;[\d_\x,\d_{\x'}]=\d_{\x''}, &&\;\x''^\m=\x^\n\pa_\n\x'^\m-\x'^\n\pa_\n\x^\m,\NO\\
%%%%%%%%%%%%%%%%%%%%%%%%%%%%%%%%%%%%%%%%%%%%%%%%%%%%%%%%%%%%%%%%%%%%%%%%%%
&\;[\d_\l,\d_{\l'}]=\d_{\l''}, &&\;\l''^a{}_b=\l'^a{}_c\l^c{}_b-\l^a{}_c\l'^c{}_b,\rule{.0cm}{.5cm}\NO\\
%%%%%%%%%%%%%%%%%%%%%%%%%%%%%%%%%%%%%%%%%%%%%%%%%%%%%%%%%%%%%%%%%%%%%%%%%%
&\;[\d_\ve,\d_\h]=\d_\s+\d_\l+\d_\th, &&\;\s=\frac12\lbar\ve\h,\quad  \l^a{}_b=-\frac12\lbar\ve\g^a{}_b\h,\quad \th=-\frac{3i}{4}\lbar\ve\g^5\h,\NO\\
%%%%%%%%%%%%%%%%%%%%%%%%%%%%%%%%%%%%%%%%%%%%%%%%%%%%%%%%%%%%%%%%%%%%%%%%%%
&\;[\d_\ve,\d_{\ve'}]=\d_\x+\d_\l+\d_\th, \qquad &&\;\x^\m=\frac12\lbar\ve'\g^\m\ve,\quad \l^a{}_b=-\frac12(\lbar\ve'\g^\n\ve)\;\o_\n{}^a{}_b,\quad \th=-\frac12(\lbar\ve'\g^\n\ve)A_\n.
%%%%%%%%%%%%%%%%%%%%%%%%%%%%%%%%%%%%%%%%%%%%%%%%%%%%%%%%%%%%%%%%%%%%%%%%%%
\end{alignat}
Notice that the composite parameters resulting from the commutator of two $\cq$-supersymmetry transformations are field dependent, which means that the structure constants of the gauge algebra are field dependent. Such algebras are often termed {\em soft algebras} (see \cite{deWit:2018dix} for a recent discussion of soft algebras and their BRST cohomology) and supergravity theories are typically based on soft algebras. The commutation relations \eqref{CS-local-algebra} form the basis for the Wess-Zumino consistency condition analysis to determine the superconformal anomalies in $\cn=1$ conformal supergravity.

\section{Classical Ward identities}
\label{CWIDs}
\setcounter{equation}{0}

We now turn to the derivation of the classical Ward identities of a local quantum field theory coupled to background $\cn=1$ conformal supergravity. These identities can be thought of as Noether's conservation laws following from the local symmetry transformations \eqref{sugra-trans}. Since these depend only on the structure of the background supergravity, the resulting Ward identities are independent of the specific field theory Lagrangian, provided the coupling of the theory to background supergravity preserves the local supergravity symmetries at the classical level. 

The classical Ward identities can be expressed in the compact form 
\be\label{Classical-WIDs}
\d_\O\mathscr{W}[e,A,\j]=0,
\ee 
where $\O=(\x,\s,\l,\th,\ve,\h)$ denotes any of the local transformations \eqref{sugra-trans} and $\mathscr{W}[e,A,\j]$ is the generating functional of connected correlation functions of local current operators associated with the background supergravity fields, namely
\be\label{currents}
\ct^\m_a=e^{-1}\frac{\d\mathscr{W}}{\d e^a_\m},
\qquad \cj^\m=e^{-1}\frac{\d\mathscr{W}}{\d A_\m},\qquad
\cs^\m=e^{-1}\frac{\d\mathscr{W}}{\d\lbar\j_\m},
\ee
where $e\equiv \det(e^a_\m)$. These definitions do not rely on a Lagrangian description of the quantum field theory, but if such a description exists, then the generating functional $\mathscr{W}[e,A,\j]$ is expressed as 
\be
\mathscr{W}[e,A,\j]=-i\log \mathscr{Z}[e,A,\j],
\ee
where $\mathscr{Z}[e,A,\j]$ is obtained from the path integral
\be
\mathscr{Z}[e,A,\j]=\int[\tx d\F] e^{iS[\F;e,A,\j]},
\ee
over the microscopic fields $\F$. At the classical level, therefore, the generating function $\mathscr{W}[e,A,\j]$ corresponds to the classical action $S[\F;e,A,\j]$, with the microscopic fields $\F$ evaluated on-shell. 

Given the definition of the current operators \eqref{currents} and the local symmetry transformations of the background supergravity fields \eqref{sugra-trans}, classical invariance of $\mathscr{W}[e,A,\j]$ leads to a conservation law -- or Ward identity -- for each local symmetry, which we will now derive. 

\subsubsection*{Diffeomorphisms}

The transformation of the generating functional under diffeomorphisms is given by
\bal
\d_\x \mathscr{W}=&\;\int d^4x\;e\big(\d_\x e^a_\m \ct^\m_a+\d_\x A_\m\cj^\m+\d_\x\lbar\j_\m \cs^\m\big)\NO\\
=&\;\int d^4x\;e\Big(\nabla_\m(\x^\n e^a_\n) \ct^\m_a-\x^\n\o_\n{}^a{}_be^b_\m\ct^\m_a+\big(\x^\n F_{\n\m}+\pa_\m(A_\n\x^\n)\big)\cj^\m\NO\\
&+\x^\n(\lbar\j_\m \overleftarrow \cd_\n-\lbar\j_\n \overleftarrow \cd_\m)\cs^\m+(\x^\n\lbar\j_\n)\overleftarrow \cd_\m \cs^\m-i\x^\n A_\n\lbar\j_\m\g^5 \cs^\m+\frac14\x^\n\o_{\n ab}\lbar\j_\m\g^{ab}\cs^\m\Big)\NO\\
=&\;\int d^4x\;e\;\x^\n\Big(-e^a_\n\nabla_\m\ct^\m_a-\nabla_\m(\lbar\j_\n  \cs^\m)+(\lbar\j_\m\overleftarrow \cd_\n)\cs^\m+F_{\n\m}\cj^\m\NO\\
&-A_\n\big(\nabla_\m\cj^\m+i\lbar\j_\m \g^5\cs^\m\big)+\o_\n{}^{ab}\big(e_{\m [a}\ct^\m_{b]}+\frac14\lbar\j_\m\g_{ab}\cs^\m\big)\Big).
\eal
Setting this quantity to zero for arbitrary $\x^\n(x)$ gives the classical diffeomorphism Ward identity 
\begin{align}
\label{diffWID}
%\boxed{
\begin{aligned}
&e^a_\m\nabla_\n\ct^\n_a+\nabla_\n(\lbar\j_\m  \cs^\n)-\lbar\j_\n\overleftarrow \cd_\m \cs^\n-F_{\m\n}\cj^\n\\
&\hspace{.5cm}+A_\m\big(\nabla_\n\cj^\n+i\lbar\j_\n \g^5\cs^\n\big)-\o_\m{}^{ab}\Big(e_{\n [a}\ct^\n_{b]}+\frac14\lbar\j_\n\g_{ab}\cs^\n\Big)=0.
\end{aligned}
%}
\end{align}
We will see shortly that the terms in the second line correspond to the classical Ward identities for $U(1)_R$ gauge transformations and local frame rotations respectively.

\subsubsection*{Weyl symmetry}

Under local Weyl rescalings the generating function transforms as
\bal
\d_\s \mathscr{W}=&\;\int d^4x\;e\big(\d_\s e^a_\m \ct^\m_a+\d_\s A_\m\cj^\m+\d_\s\lbar\j_\m \cs^\m\big)\NO\\
=&\;\int d^4x\;e\;\s\Big(e^a_\m \ct^\m_a+\frac12\lbar\j_\m \cs^\m\Big),
\eal
and, hence, the classical trace Ward identity takes the form
\be
e^a_\m\ct^\m_a+\frac12\lbar\j_\m \cs^\m=0.
\ee

\subsubsection*{$R$-symmetry}

The transformation of the generating function under $U(1)_R$ gauge transformations is given by
\bal
\d_\th \mathscr{W}=&\;\int d^4x\;e\big(\d_\th e^a_\m \ct^\m_a+\d_\th A_\m\cj^\m+\d_\th\lbar\j_\m \cs^\m\big)\NO\\
=&\;\int d^4x\;e\big(\pa_\m\th \cj^\m-i\th\lbar\j_\m\g^5 \cs^\m\big)\NO\\
=&\;\int d^4x\;e\;\th\big(-\nabla_\m \cj^\m-i\lbar\j_\m\g^5 \cs^\m\big).
\eal
Hence, the classical $R$-symmetry Ward identity takes the form
\be
\nabla_\m \cj^\m+i\lbar\j_\m\g^5 \cs^\m=0.
\ee

\subsubsection*{Local frame rotations}    

Under local frame rotations the generating function transforms according to 
\bal
\d_\l \mathscr{W}=&\;\int d^4x\;e\big(\d_\l e^a_\m \ct^\m_a+\d_\l A_\m\cj^\m+\d_\l\lbar\j_\m \cs^\m\big)\NO\\
=&\;-\int d^4x\;e\l^{ab}\Big(e_{\m[b} \ct^\m_{a]}+\frac14\lbar\j_\m\g_{ba} \cs^\m\Big).
\eal
Hence, the corresponding classical Ward identity is
\be\label{LorentzWID}
e_{\m[a} \ct^\m_{b]}+\frac14\lbar\j_\m\g_{ab} \cs^\m=0.
\ee

\subsubsection*{$\cq$-supersymmetry} 

The $\cq$-supersymmetry transformation of the generating function is
\bal
\d_\ve \mathscr{W}=&\;\int d^4x\;e\big(\d_\ve e^a_\m \ct^\m_a+\d_\ve A_\m\cj^\m+\d_\ve\lbar\j_\m \cs^\m\big)\NO\\
=&\;\int d^4x\;e\Big(-\frac12\lbar\j_\m\g^a\ve \ct^\m_a+\frac{3i}{4}\lbar\f_\m\g^5\ve\cj^\m+\lbar\ve\overleftarrow \cd_\m \cs^\m\Big)\NO\\
=&\;\int d^4x\;e\;\lbar\ve\Big(\frac12\g^a\j_\m \ct^\m_a+\frac{3i}{4}\g^5\f_\m\cj^\m- \cd_\m \cs^\m\Big).
\eal
Therefore, the classical $\cq$-supersymmetry Ward identity takes the from
\be
\cd_\m \cs^\m=\frac12\g^a\j_\m \ct^\m_a+\frac{3i}{4}\g^5\f_\m\cj^\m.
\ee

\subsubsection*{$\cs$-supersymmetry} 

Finally, the transformation of the generating function under $\cs$-supersymmetry is given by
\bal
\d_\h \mathscr{W}=&\;\int d^4x\;e\big(\d_\h e^a_\m \ct^\m_a+\d_\h A_\m\cj^\m+\d_\h\lbar\j_\m \cs^\m\big)\NO\\
=&\;\int d^4x\;e\Big(-\frac{3i}{4}\lbar\j_\m\g^5\h\cj^\m+\lbar\h\g_\m \cs^\m\Big)\NO\\
=&\;\int d^4x\;e\;\lbar\h\Big(-\frac{3i}{4}\g^5\j_\m\cj^\m+\g_\m \cs^\m\Big),
\eal
and hence the classical Ward identity for $\cs$-supersymmetry is
\be%\boxed{
\g_\m \cs^\m-\frac{3i}{4}\g^5\j_\m\cj^\m=0.%}
\ee

\section{Superconformal anomalies from the Wess-Zumino consistency conditions}
\label{WZanomalies}
\setcounter{equation}{0}

At the quantum level the generating function $\mathscr{W}$ may not be invariant under all local symmetries of background conformal supergravity, i.e.
\be
\d_\O\mathscr{W}\neq 0.
\ee
The non-invariance of the generating function of four dimensional theories can be parameterized as  
\be\label{W-anomalies-CS}
%\boxed{
\d_{\O} \mathscr{W}=\int d^4x\sqrt{-g}\;\big(\s\ca_{W}-\th\ca_{R}-\lbar\ve\ca_{Q}+\lbar\h\ca_{S}\big),%}
\ee
where $\ca_W$, $\ca_R$, $\ca_{Q}$ and $\ca_S$ are possible quantum anomalies under Weyl, $R$-symmetry, $\cq$- and $\cs$-supersymmetry transformations, respectively. Recall that the gravitational and Lorentz (frame rotation) anomalies are related by a local counterterm \cite{Bardeen:1984pm} and exist only in $4k+2$ dimensions, with $k=0,1,\ldots$. Moreover, the mixed anomaly can be moved entirely to the conservation of the $R$-current by a choice of local counterterms (that is setting $\a=0$ in 
eq.~(2.43) of \cite{Jensen:2012kj}). In this scheme diffeomorphisms remain a symmetry at the quantum level and the transformation of the generating function under all local symmetries can be parameterized as in \eqref{W-anomalies-CS}. 

Repeating the exercise of the previous section with the anomalous transformation \eqref{W-anomalies-CS} results in the same diffeomorphism and Lorentz Ward identities as those obtained respectively in \eqref{diffWID} and \eqref{LorentzWID}, but the remaining Ward identities become
\bal
\label{WardIDs}
%&e^a_\m\nabla_\n\ct^\n_a+\nabla_\n(\lbar\j_\m  \cs^\n)-\lbar\j_\n\overleftarrow \cd_\m \cs^\n-F_{\m\n}\cj^\n\NO\\
%&\hspace{2.cm}+A_\m\big(\nabla_\n\cj^\n+i\lbar\j_\n \g^5\cs^\n\big)-\o_\m{}^{ab}\Big(e_{\n [a}\ct^\n_{b]}+\frac14\lbar\j_\n\g_{ab}\cs^\n\Big)=0,\NO\\
%%%%%%%%%%%%%%%%%%%%%%%%%%%%%%%%%%%%%%%%%%%%%%%%%%%%%%%%%%%%%%%%%%%%%%%%
&e^a_\m\ct^\m_a+\frac12\lbar\j_\m \cs^\m=\ca_W,\NO\\
%%%%%%%%%%%%%%%%%%%%%%%%%%%%%%%%%%%%%%%%%%%%%%%%%%%%%%%%%%%%%%%%%%%%%%%%
%&e_{\m[a} \ct^\m_{b]}+\frac14\lbar\j_\m\g_{ab} \cs^\m=0,\NO\\
%%%%%%%%%%%%%%%%%%%%%%%%%%%%%%%%%%%%%%%%%%%%%%%%%%%%%%%%%%%%%%%%%%%%%%%%
&\nabla_\m \cj^\m+i\lbar\j_\m\g^5 \cs^\m=\ca_R,\NO\\
%%%%%%%%%%%%%%%%%%%%%%%%%%%%%%%%%%%%%%%%%%%%%%%%%%%%%%%%%%%%%%%%%%%%%%%%
&\cd_\m \cs^\m-\frac12\g^a\j_\m \ct^\m_a-\frac{3i}{4}\g^5\f_\m\cj^\m=\ca_Q,\NO\\
%%%%%%%%%%%%%%%%%%%%%%%%%%%%%%%%%%%%%%%%%%%%%%%%%%%%%%%%%%%%%%%%%%%%%%%%
&\g_\m \cs^\m-\frac{3i}{4}\g^5\j_\m\cj^\m=\ca_{S}.
\eal
The objective of this section is to determine the general form of the quantum anomalies $\ca_W$, $\ca_R$, $\ca_{Q}$ and $\ca_S$ by solving the Wess-Zumino consistency conditions \cite{Wess:1971yu} associated with the $\cn=1$ conformal supergravity algebra \eqref{CS-local-algebra}.

The Wess-Zumino consistency conditions amount to the requirement that the local symmetry algebra \eqref{CS-local-algebra} is realized when successive infinitesimal local symmetry variations $\d_\O$ (also known as Ward operators) act on the generating functional $\mathscr{W}$ and read
\be\label{WZ-condition}
[\d_\O,\d_{\O'}] \mathscr{W}=\d_{[\O,\O']}\mathscr{W},
\ee
for any pair of local symmetries $\O=(\x,\s,\l,\th,\ve,\h)$ and $\O'=(\x',\s',\l',\th',\ve',\h')$.  

In appendix \ref{WZ-proof} we determine general non trivial solution of the Wess-Zumino consistency conditions \eqref{WZ-condition} for the $\cn=1$ conformal supergravity algebra \eqref{CS-local-algebra} in the scheme where diffeomorphisms and local Lorentz transformations are non anomalous. There are two non trivial solutions related respectively to the $a$ and $c$ coefficients of the Weyl anomaly and they take the form
\bal\label{anomalies}
\ca_W=&\;\frac{c}{16\p^2}\Big(W^2-\frac{8}{3}F^2\Big)-\frac{a}{16\p^2} E+\co(\j^2),\NO\\
%%%%%%%%%%%%%%%%%%%%%%%%%%%%%%%%%%%%%%%%%%%%%%%%%%%%%%%%%%%%%%%%%%%%%%%%
\ca_R=&\;\frac{(5a-3c)}{27\p^2}\;\wt F F+\frac{(c-a)}{24\p^2}\cp,\NO\\
%%%%%%%%%%%%%%%%%%%%%%%%%%%%%%%%%%%%%%%%%%%%%%%%%%%%%%%%%%%%%%%%%%%%%%%%
\ca_Q=&\;-\frac{(5a-3c)i}{9\p^2}\wt F^{\m\n}A_\m\g^5\f_\n+\frac{(a-c)}{6\p^2}\nabla_\m\big(A_\r \wt R^{\r\s\m\n} \big)\g_{(\n}\j_{\s)}-\frac{(a-c)}{24\p^2}F_{\m\n} \wt R^{\m\n\r\s} \g_\r\j_\s+\co(\j^3),\NO\\
%%%%%%%%%%%%%%%%%%%%%%%%%%%%%%%%%%%%%%%%%%%%%%%%%%%%%%%%%%%%%%%%%%%%%%%%
\ca_{S}=&\;\frac{(5a-3c)}{6\p^2}\wt F^{\m\n}\Big(\cd_\m-\frac{2i}{3}A_\m\g^5\Big)\j_{\n}+\frac{ic}{6\p^2} F^{\m\n}\big(\g_{\m}{}^{[\s}\d_{\n}^{\r]}-\d_{\m}^{[\s}\d_{\n}^{\r]}\big)\g^5\cd_\r\j_\s\NO\\
&+\frac{3(2a-c)}{4\p^2}P_{\m\n}g^{\m[\n}\g^{\r\s]}\cd_\r\j_\s+\frac{(a-c)}{8\p^2}\Big(R^{\m\n\r\s}\g_{\m\n}-\frac12Rg_{\m\n}g^{\m[\n}\g^{\r\s]}\Big)\cd_\r\j_\s+\co(\j^3),
\eal
where $W^2$ is the square of the Weyl tensor, $E$ is the Euler density and $\cp$  is the Pontryagin density. Their expressions in terms of the Riemann tensor are  
\bal
W^2\equiv&\; W_{\m\n\r\s}W^{\m\n\r\s}=R_{\m\n\r\s}R^{\m\n\r\s}-2R_{\m\n}R^{\m\n}+\frac13R^2,\NO\\
E=&\;R_{\m\n\r\s}R^{\m\n\r\s}-4R_{\m\n}R^{\m\n}+R^2,\NO\\
\cp\equiv&\;\frac12\e^{\k\l\m\n}R_{\k\l\r\s}R_{\m\n}{}^{\r\s}=\wt R^{\m\n\r\s}R_{\m\n\r\s},
\eal
where the dual Riemann tensor is defined in analogy with the dual U(1)$_R$ fieldstrength in \eqref{dualF} as\footnote{In contrast to the Riemann tensor, $\wt R_{\m\n\r\s}$ is not symmetric under exchange of the first and second pair of indices.}
\be
\label{dualR}
\wt R_{\m\n\r\s}\equiv\frac12\e_{\m\n}{}^{\k\l}R_{\k\l\r\s}.
\ee
Moreover, $P_{\m\n}$ is the Schouten tensor defined in \eqref{Schouten} and we have introduced the shorthand notation
\be
F^2\equiv F_{\m\n}F^{\m\n},\qquad F\wt F\equiv \frac12 \e^{\m\n\r\s}F_{\m\n}F_{\r\s}.
\ee
Finally, we have used the normalization of the central charges adopted in \cite{Anselmi:1997am}, according to which the $a$ and $c$ anomaly coefficients for free chiral and vector multiplets are given respectively by 
\be
a=\frac{1}{48}(N_\c+9N_v),\qquad c=\frac{1}{24}(N_\c+3N_v).
\ee

Several comments are in order here. Firstly, we should point out that the anomalies \eqref{anomalies}, as well as the current operators defined in \eqref{currents}, are the {\em consistent} ones. The corresponding covariant quantities can be obtained by adding the appropriate Bardeen-Zumino terms \cite{Bardeen:1984pm}. Secondly, the bosonic Ward identities and anomalies we obtain reproduce well known results \cite{Anselmi:1997am} (corrected in \cite{Cassani:2013dba}), but the fermionic Ward identities and anomalies have not appeared -- at least explicitly -- in the literature before. Moreover, the Ward identities \eqref{diffWID}, \eqref{LorentzWID} and \eqref{WardIDs} derived above, including the anomalies \eqref{anomalies}, are in complete agreement with the results of \cite{Papadimitriou:2017kzw} for theories with a holographic dual that have
\be\label{a=c}
a=c=\frac{\p\ell^3}{8G_5},
\ee
where $G_5$ is the Newton constant in five dimensions and $\ell$ is the AdS$_5$ radius. Of course, such an agreement was expected since minimal $\cn=2$ gauged supergravity induces off-shell $\cn=1$ conformal supergravity on the four dimensional boundary of AdS$_5$ \cite{Balasubramanian:2000pq} and the anomalies can be computed through holographic renormalization \cite{Henningson:1998gx} (see also \cite{Nakayama:2012gu,Cassani:2013dba,An:2017ihs,An:2018roi}).  

An interesting question is whether there exists a local counterterm $\mathscr{W}_{ct}$ that removes the $\cq$-supersymmetry anomaly, i.e. such that $\d_\ve(\mathscr{W}+\mathscr{W}_{ct})=0$. Closure of the algebra requires that
\be
[\d_\ve,\d_{\ve'}](\mathscr{W}+\mathscr{W}_{ct})=(\d_\x+\d_\l+\d_\th)(\mathscr{W}+\mathscr{W}_{ct}),
\ee
with the composite parameters for the bosonic transformations given in \eqref{CS-local-algebra}. It follows that if such a counterterm exists, then it must also satisfy
\be
(\d_\x+\d_\l+\d_\th)(\mathscr{W}+\mathscr{W}_{ct})=0.
\ee
Hence, either $\mathscr{W}_{ct}$ removes also the $R$-symmetry anomaly, or it breaks diffeomorphisms and/or local frame rotations. This means that for theories with an $R$-symmetry anomaly, either $\cq$-supersymmetry or diffeomorphisms/Lorentz transformations are anomalous as well. The identification of a possible local counterterm that moves the $\cq$-anomaly to diffeomorphisms/Lorentz transformations is a problem we hope to address in future work.

\section{Anomalous supercurrent transformation under $\cq$- and $\cs$-supersymmetry}
\label{supercurrent-anomalies}
\setcounter{equation}{0}

The superconformal anomalies \eqref{anomalies} lead to anomalous transformations for the current operators under the corresponding local symmetries. In particular, the fermionic anomalies $\ca_Q$ and $\ca_S$ contribute to the transformation of the supercurrent under respectively $\cq$- and $\cs$-supersymmetry \cite{Papadimitriou:2017kzw}. When restricted to rigid symmetries of a specific background, the anomalous terms in the transformations of the currents result in a deformed superalgebra. 

The classical (non-anomalous) part of the current transformations can be deduced directly from the supergravity transformations \eqref{sugra-trans} and the definition of the currents in \eqref{currents}. For example, \eqref{sugra-trans} imply that the functional derivative with respect to the gravitino transforms according to
\bal
\d_\ve\Big(\frac{\d}{\d\lbar\j_\m}\Big)
=&\;\frac{1}{2}\g^a\ve\frac{\d}{\d e^a_\m}+\frac{i}{8}\big(4\d^{[\m}_\n\d^{\r]}_\s+i\g^5 \e^\m{}_{\n}{}^{\r}{}_\s\big) \g^\n\g^5\cd_\r\Big(\ve\frac{\d}{\d A_\s}\Big),\NO\\
\d_\h\Big(\frac{\d}{\d\lbar\j_\m}\Big)=&\;\frac{3i}{4}\g^5\h\frac{\d}{\d A_\m}.
\eal
It follows that the $\cq$- and $\cs$-supersymmetry transformations of the supercurrent are given by\footnote{An equivalent but more formal way to determine how the currents transform under the local symmetries is to utilize the symplectic structure underlying the space of couplings and local operators \cite{Papadimitriou:2016yit}. The Ward identities correspond to first class constraints on this space, generating the local symmetry transformations under the Poisson bracket. In appendix B.1 of \cite{Papadimitriou:2017kzw} this approach is used to obtain the anomalous transformation of the supercurrent under $\cq$- and $\cs$-supersymmetry in the case $a=c$.} 
\bal\label{Q-supercurrent-trans}
\d_\ve\cs^\m=&\;e^{-1}\d_\ve\Big(\frac{\d}{\d\lbar\j_\m}\Big)\mathscr{W}+e^{-1}\frac{\d}{\d\lbar\j_\m}\d_\ve\mathscr{W}\NO\\
=&\;\frac{1}{2}\g^a\ve\ct^\m_a+\frac{i}{8}\big(4\d^{[\m}_\n\d^{\r]}_\s+i\g^5 \e^\m{}_{\n}{}^{\r}{}_\s\big) \g^\n\g^5\cd_\r\Big[\ve\Big(\cj^\s+\frac{4(5a-3c)}{27\p^2}\wt F^{\s\k}A_\k\Big)\Big]\NO\\
&+\frac{(a-c)}{6\p^2}\de_{\r}\big(A_{\s}\wt R^{\s\l\r\k}\big)\d^\m_{(\k}\g_{\l)}\ve+\frac{(a-c)}{24\p^2}F_{\r\s}\wt R^{\r\s\m\n}\g_\n\ve,\\\NO\\
%%%%%%%%%%%%%%%%%%%%%%%%%%%%%%%%%%%%%%%%%%%%%%%%%%%%%%%%%%%%%%%%%%%%%%%%%%%%%%%%%%%
\d_\h\cs^\m=&\;e^{-1}\d_\h\Big(\frac{\d}{\d\lbar\j_\m}\Big)\mathscr{W}+e^{-1}\frac{\d}{\d\lbar\j_\m}\d_\h\mathscr{W}\NO\\
=&\;\frac{3i}{4}\g^5\h\Big(\cj^\m+\frac{4(5a-3c)}{27\p^2}\wt F^{\m\n}A_\n\Big)+\frac{(5a-3c)}{6\p^2}\cd_\n(\wt F^{\m\n}\h)-\frac{ic}{6\p^2}\big(\g^{[\m}{}_\r\d^{\n]}_\s-\d^{[\m}_\r\d^{\n]}_\s\big)\g^5\cd_\n(F^{\r\s}\h)\NO\\
&-\frac{3(2a-c)}{4\p^2}\cd_\n\big(P_{\r\s}g^{\r[\s}\g^{\m\n]}\h\big)-\frac{(a-c)}{8\p^2}\cd_\n\Big[\Big(R^{\m\n\r\s}\g_{\r\s}-\frac12Rg_{\r\s}g^{\r[\s}\g^{\m\n]}\Big)\h\Big].\label{S-supercurrent-trans}
\eal
Notice that these transformations coincide with those in eq.~(5.9) of \cite{Papadimitriou:2017kzw} in the special case $a=c$. 

The anomalous transformations of the supercurrent in \eqref{Q-supercurrent-trans} and \eqref{S-supercurrent-trans} are essentially a rewriting of the two fermionic Ward identities in \eqref{WardIDs} and are useful for e.g. determining the effect of the superconformal anomalies in correlation functions \cite{Katsianis:2019hhg}. They also determine the rigid superalgebra on curved backgrounds that admit Killing spinors of conformal supergravity. Namely, when the local spinor parameters $\ve$ and $\h$ are restricted to solutions $(\ve_o,\h_o)$ of the Killing spinor equation
\be
\d_{\ve_o,\h_o}\j_\m=\cd_\m\ve_o-\g_\m\h_o=0,
\ee    
on a fixed bosonic background specified by $g_{\m\n}$ and $A_\m$, the corresponding transformation of the supercurrent under {\em rigid} supersymmetry is given by\footnote{Note that $(\ve_o,\h_o)$ are $c$-number parameters, while $(\ve,\h)$ are Grassmann valued local parameters that transform non trivially under the local symmetries according to \eqref{param-trans}.}
\be
\d_{(\ve_o,\h_o)}\cs^\m=\{\lbar\cq[\ve_o,\h_o],\cs^\m\}=(\d_{\ve_o}+\d_{\h_o})\cs^\m,
\ee
where $\lbar\cq[\ve_o,\h_o]$ is the conserved supercharge associated with the Killing spinor $(\ve_o,\h_o)$ and all bosonic fields in the transformations \eqref{Q-supercurrent-trans} and \eqref{S-supercurrent-trans} are evaluated on the specific background. In \cite{Papadimitriou:2017kzw,An:2017ihs} it was shown that even though the Weyl and $R$-symmetry anomalies are numerically zero for a class of $\cn=1$ conformal supergravity  backgrounds admitting two real supercharges of opposite $R$-charge \cite{Klare:2012gn,Dumitrescu:2012ha}, the anomalous terms in the transformations of the supercurrent under rigid supersymmetry do not vanish, leading to a deformed rigid superalgebra on such backgrounds. As reviewed in the Introduction, this observation was the key to resolving the apparent tension between the field theory results of     \cite{Closset:2013vra,Closset:2014uda,Assel:2014paa} that used the classical superalgebra and the holographic computation of \cite{Genolini:2016ecx}. Besides the dependence of the supersymmetric partition function on the background, however, the transformation of the supercurrent determines also the spectrum of BPS states. The anomalous transformation of the supercurrent results in a shifted spectrum \cite{Papadimitriou:2017kzw,An:2017ihs}.

Although it may not be desirable -- or even possible -- to eliminate the $\cq$-anomaly by a local counterterm that breaks diffeomorphisms and/or local Lorentz transformations as discussed in the previous section, it is plausible that the anomaly in {\em rigid} supersymmetry may be removed by a local counterterm that breaks certain (large) diffeomorphisms, but preserves the underlying structure of the supersymmetric background. An example of a somewhat analogous situation was discussed in \cite{Imbimbo:2014pla}, where supersymmetric gauge theories in three dimensions with both Maxwell and Chern-Simons terms coupled to background topological gravity were considered. The partition function of such theories on Seifert manifolds, which admit two supercharges of opposite $R$-charge, can be computed via supersymmetric localization and depends explicitly on the Seifert structure modulus $b$. In principle, this dependence could be explained by the presence of a framing anomaly, which implies that the partition function does indeed depend on the metric at the quantum level. The puzzle, however, is that for Seifert manifolds specifically the framing anomaly is numerically zero. The authors of \cite{Imbimbo:2014pla} resolve the puzzle by arguing that in order to make the quantum theory invariant under Seifert reparameterizations {\em and} Seifert-topological (i.e. independent of metric deformations that preserve the Seifert structure) they need to add a local but non fully covariant counterterm. This counterterm does depend on the Seifert structure modulus $b$, which resolves the paradox. 

It is plausible that similarly the rigid supersymmetry anomaly in four dimensions can be removed by a local counterterm that breaks those large diffeomorphisms that are not compatible with the structure of the supersymmetric background. In fact, for holographic theories (i.e. $a=c$ at large $N$) defined on trivial circle fibrations over Seifert manifolds such a local counterterm was found in \cite{Genolini:2016ecx}. It would be interesting to generalize this counterterm to non holographic theories with arbitrary $a$ and $c$ using the general form of the fermionic anomalies we obtained in this paper.

\section{Concluding remarks}
\label{conclusion}
\setcounter{equation}{0}

In this paper we determined the local symmetry algebra of $\cn=1$ off-shell conformal supergravity in four dimensions and obtained the general form of the superconformal anomalies by solving the associated Wess-Zumino consistency conditions. To the best of our knowledge, the explicit form of the fermionic Ward identities and their anomalies have not appeared in the literature before. We find that the divergence of the supercurrent, which is associated with $\cq$-supersymmetry, is anomalous whenever $R$-symmetry is anomalous. This anomaly cannot be removed by a local counterterm without breaking diffeomorphisms and/or local Lorentz transformations.

Several open questions remain. Our result that $\cq$-supersymmetry is anomalous in any theory with an anomalous $R$-symmetry does not seem to depend on the specific supergravity theory we used in this paper. Indeed, we expect this result to hold in non-superconformal theories with an anomalous $R$-symmetry as well. This expectation is supported by the recent analysis of \cite{An:2019zok}. Another interesting question is whether there exists a local counterterm that eliminates the $\cq$-supersymmetry anomaly. As we saw in section \ref{WZanomalies}, such a counterterm would necessarily break diffeomorphisms and/or local Lorentz rotations. The related question for {\em rigid} supersymmetry on backgrounds that admit Killing spinors is relevant for the validity of supersymmetric localization computations on four-manifolds. An important example is the computation of generalized supersymmetric indices that count the microstates of supersymmetric AdS$_5$ black holes \cite{Kinney:2005ej,Cabo-Bizet:2018ehj,Choi:2018hmj,Benini:2018ywd,Honda:2019cio}. We hope to return to these questions in future work.

\section*{Acknowledgments}

I would like to thank Benjamin Assel, Roberto Auzzi, Loriano Bonora, Davide Cassani, Cyril Closset, Camillo Imbimbo, Heeyeon Kim, Zohar Komargodski, Dario Martelli, Sameer Murthy and Dario Rosa for illuminating discussions and email correspondence. I am also grateful to the University of Southampton, King's College London and the International Center for Theoretical Physics in Trieste for hospitality and partial financial support during the completion of this work.

\appendix

\renewcommand{\theequation}{\Alph{section}.\arabic{equation}}

\setcounter{section}{0}

\section*{Appendix}
\setcounter{section}{0}

\section{Spinor conventions and identities}
\label{conventions}
\setcounter{equation}{0}	

Throughout this paper we follow the conventions of \cite{Freedman:2012zz}. In particular, the tangent space metric is $\h=\diag(-1,1,1,1)$ and the Levi-Civita symbol $\ve_{\m\n\r\s}=\pm 1$ satisfies $\ve_{0123}=1$. The Levi-Civita tensor is defined as usual as $\e_{\m\n\r\s}=\sqrt{-g}\;\ve_{\m\n\r\s}=e\;\ve_{\m\n\r\s}$. Moreover, the chirality matrix in four dimensions is given by
\be
\g^5=i\g_0\g_1\g_2\g_3,
\ee
and we define the antisymmetrized products of gamma matrices as
\be
\g^{\m_1\m_2\ldots\m_n}\equiv\g^{[\m_1}\g^{\m_2}\cdots\g^{\m_n]},
\ee
where antisymmetrization is done with weight one.

In the conventions we use here the gravitino $\j_\m$ is a Majorana spinor (see section 3.3 of \cite{Freedman:2012zz} for the definition) and we make extensive use of the spinor bilinear identity in four dimensions
\be
\lbar\l\g^{\m_1}\g^{\m_2}\cdots\g^{\m_p}\c=(-1)^p\lbar\c\g^{\m_p}\cdots\g^{\m_2}\g^{\m_1}\l,\qquad \text{[eq.~(3.53) in \cite{Freedman:2012zz}]}.
\ee
For Majorana fermions we also have that 
\be
(\lbar\c\g_{\m_1\ldots\m_r}\l)^*=\lbar\c\g_{\m_1\ldots\m_r}\l,\qquad \text{[eq.~(3.82) in \cite{Freedman:2012zz}]}.
\ee

It is convenient to collect several identities involving antisymmetrized products of gamma matrices in $d$ dimensions, most of which can be found in section 3 of \cite{Freedman:2012zz}:  
\bal\label{d-gamma-ids} 
\g^{\m\n\r}&=\frac12\{\g^\m,\g^{\n\r}\},\NO\\
%%%%%%%%%%%%%%%%%%%%%%%%%%%%%%%%%%%%%%%%%%%%%%%%%%%%%%%%%%%%%%%%%%%%%%%
\g^{\m\n\r\s}&=\frac12[\g^\m,\g^{\n\r\s}],\NO\\
%%%%%%%%%%%%%%%%%%%%%%%%%%%%%%%%%%%%%%%%%%%%%%%%%%%%%%%%%%%%%%%%%%%%%%%%%%%%%%
\g^{\m\n}\g_{\r\s}&=\g^{\m\n}{}_{\r\s}+4\g^{[\m}{}_{[\s}\d^{\n]}{}_{\r]}+2\d^{[\m}{}_{[\s}\d^{\n]}{}_{\r]},\NO\\
%%%%%%%%%%%%%%%%%%%%%%%%%%%%%%%%%%%%%%%%%%%%%%%%%%%%%%%%%%%%%%%%%%%%%%%%%
\g_{\m}\g^{\n_1\ldots\n_p}&=\g_\m{}^{\n_1\ldots\n_p}+p\d ^{[\n_1}_{\m}\g^{\n_2\ldots\n_p]},\NO\\
%%%%%%%%%%%%%%%%%%%%%%%%%%%%%%%%%%%%%%%%%%%%%%%%%%%%%%%%%%%%%%%%%%%%%%%%%
\g^{\n_1\ldots\n_p}\g_{\m}&=\g^{\n_1\ldots\n_p}{}_{\m}+p\g^{[\n_1\ldots\n_{p-1}}\d ^{\n_p]}_{\m},\NO\\
%%%%%%%%%%%%%%%%%%%%%%%%%%%%%%%%%%%%%%%%%%%%%%%%%%%%%%%%%%%%%%%%%%%%%%%%%
\g^{\m\n\r}\g_{\s\t}&=\g^{\m\n\r}{}_{\s\t}+6\g^{[\m\n}{}_{[\t}\d ^{\r]}{}_{\s]}+6\g^{[\m}\d^\n{}_{[\t}\d^{\r]}{}_{\s]},\NO\\
%%%%%%%%%%%%%%%%%%%%%%%%%%%%%%%%%%%%%%%%%%%%%%%%%%%%%%%%%%%%%%%%%%%%%%%%%%%%%%
\g^{\m\n\r\s}\g_{\t\l}&=\g^{\m\n\r\s}{}_{\t\l}+8\g^{[\m\n\r}{}_{[\l}\d^{\s]}{}_{\t]}+12\g^{[\m\n}\d^\r{}_{[\l}\d^{\s]}{}_{\t]},\NO\\
%%%%%%%%%%%%%%%%%%%%%%%%%%%%%%%%%%%%%%%%%%%%%%%%%%%%%%%%%%%%%%%%%%%%%%%%%%%%%%
\g^{\m\n\r}\g_{\s\t\l}&=\g^{\m\n\r}{}_{\s\t\l}+9\g^{[\m\n}{}_{[\t\l}\d^{\r]}{}_{\s]}+18\g^{[\m}{}_{[\l}\d^\n{}_{\t}\d^{\r]}{}_{\s]}+6\d^{[\m}{}_{[\l}\d^\n{}_{\t}\d^{\r]}{}_{\s]},\NO\\
%%%%%%%%%%%%%%%%%%%%%%%%%%%%%%%%%%%%%%%%%%%%%%%%%%%%%%%%%%%%%%%%%%%%%%%%%%%%%%
\g^{\m_1\ldots\m_r\n_1\ldots\n_s}\g_{\n_s\ldots\n_1}&=\frac{(d-r)!}{(d-r-s)!}\g^{\m_1\ldots\m_r},\NO\\
%%%%%%%%%%%%%%%%%%%%%%%%%%%%%%%%%%%%%%%%%%%%%%%%%%%%%%%%%%%%%%%%%%%%%%%%%%
\g^{\m\r}\g_{\r\n}&=(d-2)\g^{\m}{}_\n+(d-1)\d^\m_\n,\NO\\
%%%%%%%%%%%%%%%%%%%%%%%%%%%%%%%%%%%%%%%%%%%%%%%%%%%%%%%%%%%%%%%%%%%%%%%%%%
\g^{\m\n\r}\g_{\r\s}&=(d-3)\g^{\m\n}{}_\s+2(d-2)\g^{[\m}\d^{\n]}{}_\s,\NO\\
%%%%%%%%%%%%%%%%%%%%%%%%%%%%%%%%%%%%%%%%%%%%%%%%%%%%%%%%%%%%%%%%%%%%%%%%%%
\g_{\m\n}\g^{\n\r\s}&=(d-3)\g_{\m}{}^{\r\s}+2(d-2)\d^{[\r}_\m\g^{\s]},\NO\\
%%%%%%%%%%%%%%%%%%%%%%%%%%%%%%%%%%%%%%%%%%%%%%%%%%%%%%%%%%%%%%%%%%%%%%%%%%%%%%
\g^{\m\n\l}\g_{\l\r\s}&=(d-4)\g^{\m\n}{}_{\r\s}+4(d-3)\g^{[\m}{}_{[\s}\d^{\n]}{}_{\r]}+2(d-2)\d^{[\m}{}_{[\s}\d^{\n]}{}_{\r]},\NO\\
%%%%%%%%%%%%%%%%%%%%%%%%%%%%%%%%%%%%%%%%%%%%%%%%%%%%%%%%%%%%%%%%%%%%%%%%%%%%%%%%%%
\g_{\m\r}\g^{\r\s\t}\g_{\t\n}&=(d-4)^2\g_\m{}^\s{}_\n+(d-4)(d-3)\(\g_\m\d_\n^\s-\g^\s g_{\m\n}\)\NO\\
&\hskip0.5cm+(d-3)(d-2)\d_\m^\s\g_\n-(d-3)\g^\s\g_{\m\n},\NO\\
%%%%%%%%%%%%%%%%%%%%%%%%%%%%%%%%%%%%%%%%%%%%%%%%%%%%%%%%%%%%%%%%%%%%%%%%%%%
\g_\r \g^{\m_1\m_2\ldots \m_p}\g^\r&=(-1)^p(d-2p)\g^{\m_1\m_2\ldots \m_p}.
\eal 

In particular, the following identities apply specifically to $d=4$ and are used extensively:
\bal\label{d=4-gamma-ids}
&\g^\r\g^\m\g^\s+\g^\s\g^\m\g^\r=2(g^{\m\r}\g^\s+g^{\m\s}\g^\r-g^{\r\s}\g^\m),\NO\\
&\g^\r\g^\m\g^\s-\g^\s\g^\m\g^\r=2\g^{\r\m\s},\NO\\
&\g^\r\g^\m\g^\s=g^{\m\r}\g^\s+g^{\m\s}\g^\r-g^{\r\s}\g^\m+\g^{\r\m\s}\NO\\
&\g^{\m\r}\g^\s=\g^{\m\r\s}+\g^\m g^{\r\s}-\g^\r g^{\m\s}\NO\\
&\g^{\m\r\s}=i\e^{\m\r\s\n}\g_\n\g^5,\NO\\
&\g^{\m\n}=\frac{i}{2}\e^{\m\n\r\s}\g_{\r\s}\g^5.
\eal
Finally, the following three identities in four dimensions help compare the superconformal Ward identities \eqref{WardIDs} and the anomalies \eqref{anomalies} with the corresponding results obtained in \cite{Papadimitriou:2017kzw}:
\bal
(\g_{\m\n}-2g_{\m\n})\g^{\n\k\l}=&\;4\d_\m{}^{[\k}\g^{\l]}-\g_\m{}^{\k\l}\NO\\
=&\;4\d_\m{}^{[\k}\g^{\l]}-i\;\e_\m{}^{\n\r\s}\g_{\s}\g_5,\NO\\
%%%%%%%%%%%%%%%%%%%%%%%%%%%%%%%%%%%%%%%%%%%%%%%%%%%%%%%%%%%%%%%%%%%%%%%
\g^\m\g^{\n\k\l}=&\;\g^{\m\n\k\l}+3g^{\m[\n}\g^{\k\l]},\NO\\
%%%%%%%%%%%%%%%%%%%%%%%%%%%%%%%%%%%%%%%%%%%%%%%%%%%%%%%%%%%%%%%%%%%%%%%
\(2\g^{\n\k}\g^\m-3\g^{\n\k\m} \)\g_{\m\r\s}=&\;2\g^{\n\k}\g^\m\g_{\m\r\s}-3\g^{\n\k\m}\g_{\m\r\s}\NO\\=&\; 4\g^{\n\k}\g_{\r\s}-12\big(\g^{[\n}{}_{[\s}\d^{\k]}{}_{\r]}+\d^{[\n}{}_{[\s}\d^{\k]}{}_{\r]}\big)\NO\\
=&\;4\big(\g^{\n\k}{}_{\r\s}+\g^{[\n}{}_{[\s}\d^{\k]}{}_{\r]}-\d^{[\n}{}_{[\s}\d^{\k]}{}_{\r]}\big).
\eal

\section{Solving the Wess-Zumino consistency conditions}
\label{WZ-proof}
\setcounter{equation}{0}

In this appendix we provide the details of the proof that the superconformal anomalies \eqref{anomalies} satisfy the Wess-Zumino consistency conditions \eqref{WZ-condition} associated with the local symmetry algebra \eqref{CS-local-algebra} of $\cn=1$ conformal supergravity. Only a subset of the algebra relations need be checked explicitly since all commutators between any two non-anomalous symmetries are trivially satisfied. Moreover, the Wess-Zumino conditions for purely bosonic symmetries are known to hold \cite{Bonora:1985cq} and are straightforward to check. We shall therefore focus on the commutation relations involving at least one fermionic symmetry transformation, except for the four commutators 
\be
[\d_\x,\d_\ve]\mathscr{W}=0,\qquad [\d_\x,\d_\h]\mathscr{W}=0,\qquad [\d_\l,\d_\ve]\mathscr{W}=0,\qquad [\d_\l,\d_\h]\mathscr{W}=0,
\ee
which hold trivially. All computations in this appendix assume either a compact spacetime manifold or that the fields go to zero at infinity so that total derivative terms can be dropped. Moreover, we only keep the leading non trivial terms in the gravitino $\j_\m$.

\begin{flushleft}
\begin{tabular}{l}
\hline\hline
\raisebox{0.1cm}{\mbox{$[\d_\s,\d_\ve]\mathscr{W}=0$ and $[\d_\s,\d_\h]\mathscr{W}=0$:\rule{0cm}{.5cm}\hspace{10.7cm}}}\\
\hline\hline
\end{tabular}
\end{flushleft}

Let us first consider the two commutation relations
\be
[\d_\s,\d_\ve]\mathscr{W},\qquad [\d_\s,\d_\h]\mathscr{W}.
\ee
Taking into account the Weyl transformation of the supersymmetry parameters in \eqref{param-trans}, it is straightforward to check that both the $\cq$- and $\cs$-supersymmetry anomalies are Weyl invariant, i.e.
\be
\d_\s\d_\ve\mathscr{W}=0,\qquad \d_\s\d_\h\mathscr{W}=0.
\ee
Moreover, the $\co(\j^2)$ terms in the Weyl anomaly in \eqref{anomalies} ensure that the Weyl anomaly {\em density} is invariant under both $\cq$- and $\cs$-supersymmetry, i.e.\footnote{For the case $a=c$ the $\co(\j^2)$ terms in the Weyl anomaly can be found in \cite{Papadimitriou:2017kzw}. For generic $a$ and $c$ the conditions \eqref{Weyl-anomaly-invariance} can be used to derive the fermionic terms in the Weyl anomaly.}     
\be\label{Weyl-anomaly-invariance}
\d_\ve(e\;\ca_W)=0,\qquad \d_\h(e\;\ca_W)=0.
\ee
Combining these results we conclude that these two commutators satisfy the Wess-Zumino consistency conditions compatible with the local symmetry algebra, namely
\be
[\d_\s,\d_\ve]\mathscr{W}=0,\qquad [\d_\s,\d_\h]\mathscr{W}=0.
\ee

\begin{flushleft}
\begin{tabular}{l}
\hline\hline
\raisebox{0.1cm}{\mbox{$[\d_\ve,\d_\th]\mathscr{W}=0$:\rule{0cm}{.5cm}\hspace{13.8cm}}}\\
\hline\hline
\end{tabular}
\end{flushleft}

The only remaining commutation relations involving a bosonic symmetry are those between local gauge transformations and either $\cq$- or $\cs$-supersymmetry transformations. Staring with $\cq$-supersymmetry, the anomalies in \eqref{anomalies} determine 
\bal
&\d_{\ve}\d_{\th} \mathscr{W}=-\d_{\ve}\int d^4x\;e\;\th\ca_{R}\NO\\
=&\;-\frac{(5a-3c)}{54\p^2}\int d^4 x\;e\;\th\;\e^{\m\n\r\s}\d_{\ve}(F_{\m\n}F_{\r\s})+\frac{(c-a)}{48\p^2}\int d^4 x\;e\;\th\;\e^{\k\l\m\n}\d_{\ve}(R^\r{}_{\s\k\l}R^{\s}{}_{\r\m\n})\NO\\
=&\;-\frac{2(5a-3c)}{27\p^2}\int d^4 x\;e\;\th\;\e^{\m\n\r\s}F_{\m\n}\pa_\r\d_{\ve}A_\s+\frac{(c-a)}{12\p^2}\int d^4 x\;e\;\th\;\e^{\k\l\m\n}R^\r{}_{\s\k\l}\de_\m\d_{\ve}\G^{\s}_{\r\n}\NO\\
=&\;\frac{(5a-3c)i}{18\p^2}\int d^4 x\;e\;\pa_\r\th\;\e^{\m\n\r\s}F_{\m\n}\lbar\ve\g^5\f_\s-\frac{(c-a)}{12\p^2}\int d^4 x\;e\;\pa_\m\th\;\e^{\k\l\m\n}R^{\r\s}{}_{\k\l}\de_\r\d_\ve g_{\s\n},\NO\\
=&\;\frac{(5a-3c)i}{18\p^2}\int d^4 x\;e\;\pa_\r\th\;\e^{\m\n\r\s}F_{\m\n}\lbar\ve\g^5\f_\s+\frac{(c-a)}{12\p^2}\int d^4 x\;e\;\nabla_\r\big(\pa_\m\th\;\e^{\k\l\m\n}R^{\r\s}{}_{\k\l}\big)\lbar\ve\g_{(\s}\j_{\n)},\\\NO\\
%%%%%%%%%%%%%%%%%%%%%%%%%%%%%%%%%%%%%%%%%%%%%%%%%%%%%%%%%%%%%%%%%%%%%%%%
%%%%%%%%%%%%%%%%%%%%%%%%%%%%%%%%%%%%%%%%%%%%%%%%%%%%%%%%%%%%%%%%%%%%%%%%
&\d_{\th} \d_{\ve}\mathscr{W}=-\d_{\th}\int d^4x\;e\;\lbar\ve\ca_{Q}\NO\\
=&\;\frac{(5a-3c)i}{18\p^2}\int d^4 x\;e\;\pa_\r\th\;\e^{\m\n\r\s}F_{\m\n}\lbar\ve\g^5\f_\s+\frac{(c-a)}{12\p^2}\int d^4 x\;e\;\e^{\m\n\r\s}\nabla_\k\big(\pa_\r\th\; R^{\k\l}{}_{\m\n} \big)\lbar\ve\g_{(\l}\j_{\s)}.
\eal
Hence, 
\be
[\d_\ve,\d_\th]\mathscr{W}=0,
\ee
as required by the Wess-Zumino conditions.

\begin{flushleft}
\begin{tabular}{l}
\hline\hline
\raisebox{0.1cm}{\mbox{$[\d_\h,\d_\th]\mathscr{W}=0$:\rule{0cm}{.5cm}\hspace{13.8cm}}}\\
\hline\hline
\end{tabular}
\end{flushleft}

For $\cs$-supersymmetry we have similarly 
\bal
\d_{\h}\d_{\th} \mathscr{W}=&\;-\d_{\h}\int d^4x\;e\;\th\ca_{R}\\
=&\;-\frac{(5a-3c)}{54\p^2}\int d^4 x\;e\;\th\;\e^{\m\n\r\s}\d_{\h}(F_{\m\n}F_{\r\s})\NO\\
=&\;-\frac{2(5a-3c)}{27\p^2}\int d^4 x\;e\;\th\;\e^{\m\n\r\s}F_{\m\n}\pa_\r\d_{\h}A_\s\NO\\
=&\;-\frac{(5a-3c)i}{18\p^2}\int d^4 x\;e\;\pa_\r\th\;\e^{\m\n\r\s}F_{\m\n}\lbar\h\g^5\j_\s,\\\NO\\
%%%%%%%%%%%%%%%%%%%%%%%%%%%%%%%%%%%%%%%%%%%%%%%%%%%%%%%%%%%%%%%%%%%%%%%%
%%%%%%%%%%%%%%%%%%%%%%%%%%%%%%%%%%%%%%%%%%%%%%%%%%%%%%%%%%%%%%%%%%%%%%%%
\d_{\th} \d_{\h}\mathscr{W}=&\;\d_{\th}\int d^4x\;e\;\lbar\h\ca_{S}\NO\\
=&\;-\frac{(5a-3c)i}{18\p^2}\int d^4 x\;e\;\pa_\r\th\;\e^{\m\n\r\s}F_{\m\n}\lbar\h\g^5\j_\s,
\eal
and hence,
\be
[\d_\h,\d_\th]\mathscr{W}=0.
\ee

\begin{flushleft}
\begin{tabular}{l}
\hline\hline
\raisebox{0.1cm}{\mbox{$[\d_\ve,\d_{\ve'}]\mathscr{W}=\d_\th \mathscr{W}$, with $\th=-\frac12(\lbar\ve'\g^\l\ve)A_\l$:\rule{0cm}{.5cm}\hspace{9.17cm}}}\\
\hline\hline
\end{tabular}
\end{flushleft}

There remain only the four commutators among fermionic symmetries. Applying two successive $\cq$-supersymmetry transformations on the generating function $\mathscr{W}$ gives 
\bal
\d_{\ve'} \d_{\ve}\mathscr{W}=&\;-\d_{\ve'}\int d^4x\;e\;\lbar\ve\ca_{Q}\NO\\
=&\;\frac{(5a-3c)}{18\p^2}i\int d^4 x\;e\;\e^{\m\n\r\s}F_{\m\n}A_\r\lbar\ve\g^5\d_{\ve'}\f_\s+\frac{(c-a)}{12\p^2}\int d^4x\;e\;\e^{\m\n\r\s}\nabla_\k\big(A_\r R^{\k\l}{}_{\m\n} \big)\lbar\ve\g_{(\l}\d_{\ve'}\j_{\s)}\NO\\
&+\frac{(a-c)}{48\p^2}\int d^4x\;e\;\e^{\m\n\r\s}F_{\r\s} R^{\k\l}{}_{\m\n} \lbar\ve\g_\k\d_{\ve'}\j_\l\NO\\
=&\;\frac{(5a-3c)}{18\p^2}i\int d^4 x\;e\;\e^{\m\n\r\s}F_{\m\n}A_\r\lbar\ve\g^5\Big(\frac12P_{\s\l}+\frac{i}{3}F_{\s\l}\g^5-\frac{1}{12}\e_{\s\l}{}^{\k\t}F_{\k\t}\Big)\g^\l\ve'\NO\\
&-\frac{(c-a)}{12\p^2}\int d^4x\;e\;\e^{\m\n\r\s}A_\r R^{\k\l}{}_{\m\n} \nabla_\k(\lbar\ve\g_{(\l}\cd_{\s)}\ve')\NO\\
&+\frac{(a-c)}{48\p^2}\int d^4x\;e\;\e^{\m\n\r\s}F_{\r\s} R^{\k\l}{}_{\m\n} \lbar\ve\g_\k \cd_\l\ve',
\eal
and hence
\bal
[\d_\ve,\d_{\ve'}]\mathscr{W}%=&\;-\frac{(5a-3c)}{27\p^2}\int d^4 x\;e\;\e^{\m\n\r\s}F_{\m\n}A_\r F_{\s\l}\lbar\ve'\g^\l\ve\NO\\
%&-\frac{(c-a)}{12\p^2}\int d^4x\;e\;\e^{\m\n\r\s}A_\r R^{\k\l}{}_{\m\n} \nabla_\k\nabla_{(\s}(\lbar\ve'\g_{\l)}\ve)\NO\\
%&+\frac{(a-c)}{48\p^2}\int d^4x\;e\;\e^{\m\n\r\s}F_{\r\s} R^{\k\l}{}_{\m\n} \de_\l(\lbar\ve'\g_\k \ve)\NO\\
=&\;-\frac{(5a-3c)}{27\p^2}\int d^4 x\;e\;\e^{\m\n\r\s}F_{\m\n}A_\r F_{\s\l}\lbar\ve'\g^\l\ve\NO\\
&-\frac{(c-a)}{48\p^2}\int d^4x\;e\;\e^{\m\n\r\s}A_\r R^{\k\l}{}_{\m\n} R_{\k\l\s\t}(\lbar\ve'\g^{\t}\ve)\NO\\
&-\frac{(c-a)}{24\p^2}\int d^4x\;e\;\e^{\m\n\r\s}A_\r R^{\k\l}{}_{\m\n} \nabla_\k\nabla_{\s}(\lbar\ve'\g_{\l}\ve)\NO\\
&+\frac{(a-c)}{48\p^2}\int d^4x\;e\;\e^{\m\n\r\s}F_{\r\s} R^{\k\l}{}_{\m\n} \de_\l(\lbar\ve'\g_\k \ve).
\eal
The last two terms can be rearranged as
\bal
&\hspace{-2.0cm}\e^{\m\n\r\s}A_\r R^{\k\l}{}_{\m\n}\nabla_\k\nabla_\s(\lbar\ve'\g_\l\ve)-\frac12\e^{\m\n\r\s}F_{\r\s} R^{\k\l}{}_{\m\n}\nabla_\k(\lbar\ve'\g_\l\ve)\NO\\
=&\;\e^{\m\n\r\s}A_\r R^{\k}{}_{\l\m\n}[\nabla_\k,\nabla_\s](\lbar\ve'\g^\l\ve)+\de_\s\big(\e^{\m\n\r\s}A_\r R^{\k\l}{}_{\m\n}\nabla_\k(\lbar\ve'\g_\l\ve)\big)\NO\\
=&\;\de_\s\big(\e^{\m\n\r\s}A_\r R^{\k\l}{}_{\m\n}\nabla_\k(\lbar\ve'\g_\l\ve)\big)+\e^{\m\n\r\s}A_\r R^{\k\l}{}_{\m\n}R_{\k\s\l\t}(\lbar\ve'\g^\t\ve)\NO\\
=&\;\de_\s\big(\e^{\m\n\r\s}A_\r R^{\k\l}{}_{\m\n}\nabla_\k(\lbar\ve'\g_\l\ve)\big)+\frac12\e^{\m\n\r\s}A_\r R^{\k\l}{}_{\m\n}R_{\k\l\s\t}(\lbar\ve'\g^\t\ve),
\eal
so that
\bal
[\d_\ve,\d_{\ve'}]\mathscr{W}=&\;-\frac{(5a-3c)}{27\p^2}\int d^4 x\;e\;\e^{\m\n\r\s}F_{\m\n}A_\r F_{\s\l}\lbar\ve'\g^\l\ve\NO\\
&-\frac{(c-a)}{24\p^2}\int d^4x\;e\;\e^{\m\n\r\s}A_\r R^{\k\l}{}_{\m\n} R_{\k\l\s\t}(\lbar\ve'\g^{\t}\ve).
\eal
Moreover, the fact that $F_{[\l\s}F_{\m\n}A_{\r]}=0$ and $R_{\k\l[\m\n}R^{\k\l}{}_{\s\t}A_{\r]}=0$ in four dimensions leads to the two identities  
\bal\label{5-antisym-ids}
\e^{\m\n\r\s}F_{\m\n}F_{\s\l}A_\r=&\;-\frac14\e^{\m\n\r\s}F_{\m\n}F_{\r\s}A_\l,\NO\\
\e^{\m\n\r\s}R_{\m\n\k\l}R_{\s\t}{}^{\k\l}A_\r=&\;-\frac14\e^{\m\n\r\s}R_{\m\n\k\l}R_{\r\s}{}^{\k\l}A_\t.
\eal
Therefore, we finally get 
\bal
[\d_\ve,\d_{\ve'}]\mathscr{W}=&\;\frac{(5a-3c)}{108\p^2}\int d^4 x\;e\;\e^{\m\n\r\s}F_{\m\n}F_{\r\s}(A_\l\lbar\ve'\g^\l\ve)+\frac{(c-a)}{96\p^2}\int d^4x\;e\;\e^{\m\n\r\s}R^{\k\l}{}_{\m\n}R_{\k\l\r\s} (A_\t\lbar\ve'\g^\t\ve)\NO\\
=&\;-\int d^4 x\;e\;\th\ca_R,
\eal
with
\be
\th=-\frac12(\lbar\ve'\g^\l\ve)A_\l,
\ee
as required by the Wess-Zumino consistency conditions.

\begin{flushleft}
\begin{tabular}{l}
\hline\hline
\raisebox{0.1cm}{\mbox{$[\d_\h,\d_{\h'}]\mathscr{W}=0$:\rule{0cm}{.5cm}\hspace{13.69cm}}}\\
\hline\hline
\end{tabular}
\end{flushleft}

Two successive $\cs$-supersymmetry transformations on the generating function $\mathscr{W}$ give
\bal\label{S^2}
\d_{\h'} \d_{\h}\mathscr{W}=&\;-\frac{(5a-3c)i}{18\p^2}\int d^4x\;e\;\e^{\m\n\r\s}F_{\m\n}A_\r\lbar\h\g^5\d_{\h'}\j_\s+\frac{(5a-3c)}{12\p^2}\int d^4x\;e\;\e_{\m\n}{}^{\r\s}F^{\m\n}\lbar\h \cd_\r\d_{\h'}\j_{\s}\NO\\
&+\frac{ic}{6\p^2}\int d^4x\;e\; F^{\m\n}\lbar\h\big(\g_{\m}{}^{[\s}\d_{\n}^{\r]}-\d_{\m}^{[\s}\d_{\n}^{\r]}\big)\g^5\cd_\r\d_{\h'}\j_\s+\frac{3(2a-c)}{4\p^2}\int d^4x\;e\;P_{\m\n}g^{\m[\n}\lbar\h\g^{\r\s]}\cd_\r\d_{\h'}\j_\s\NO\\
&+\frac{(a-c)}{8\p^2}\int d^4x\;e\;\lbar\h\Big(R^{\m\n\r\s}\g_{\m\n}-\frac12Rg_{\m\n}g^{\m[\n}\g^{\r\s]}\Big)\cd_\r\d_{\h'}\j_\s\NO\\
=&\;\frac{(5a-3c)i}{18\p^2}\int d^4x\;e\;\e^{\m\n\r\s}F_{\m\n}A_\r\lbar\h\g^5\g_\s\h'-\frac{(5a-3c)}{12\p^2}\int d^4x\;e\;\e_{\m\n}{}^{\r\s}F^{\m\n}\lbar\h \g_\s \cd_\r\h'\NO\\
&-\frac{ic}{6\p^2}\int d^4x\;e\; F^{\m\n}\lbar\h\big(\g_{\m}{}^{[\s}\d_{\n}^{\r]}-\d_{\m}^{[\s}\d_{\n}^{\r]}\big)\g^5\g_\s \cd_\r\h'-\frac{3(2a-c)}{4\p^2}\int d^4x\;e\;P_{\m\n}g^{\m[\n}\lbar\h\g^{\r\s]}\g_\s \cd_\r\h'\NO\\
&-\frac{(a-c)}{8\p^2}\int d^4x\;e\;\lbar\h\Big(R^{\m\n\r\s}\g_{\m\n}-\frac12Rg_{\m\n}g^{\m[\n}\g^{\r\s]}\Big)\g_\s \cd_\r\h'.
\eal
We will now show that each of these terms vanishes once the corresponding expression with $\h$ and $\h'$ interchanged is subtracted.

The first term in the second equality in \eqref{S^2} vanishes trivially since
\be
\lbar\h\g^5\g_\s\h'=\lbar\h'\g^5\g_\s\h.
\ee
For the second term we have
\bal
\lbar\h\g_\s \cd_\r\h'-\lbar\h'\g_\s \cd_\r\h=\de_\r(\lbar\h\g_\s\h').
\eal
Integrating by parts and using the Bianchi identity $\e^{\m\n\r\s}\pa_\r F_{\m\n}=0$ we find that the second term vanishes as well. 

The third term can be simplified as
\bal
&\hskip-2.cmF^{\n\k}\lbar\h\big(\g_{\n}{}^{[\s}\d_{\k}{}^{\r]}-\d_{\n}{}^{[\s}\d_{\k}{}^{\r]}\big)\g^5\g_\s \cd_\r\h'\NO\\
=&\;-\frac12F^{\n\k}\lbar\h\big(\g_{\n}{}^{\s}g_{\k}^{\r}-\g_{\n}{}^{\r}g_{\k}^{\s}-2g_{\n}^{\s}g_{\k}^{\r}\big)\g_\s\g^5 \cd_\r\h'\NO\\
=&\;-\frac12F^{\n\k}\lbar\h\big(3\g_{\n}g_{\k}^{\r}-\g_{\n}{}^{\r}{}_\k-\g_\n g^\r_\k+\g^\r \cancelto{0}{g_{\n\k}}-2g_{\k}^{\r}\g_\n\big)\g^5 \cd_\r\h'\NO\\
=&\;-\frac12F^{\n\k}\lbar\h\g^{\r}{}_{\n\k}\g^5 \cd_\r\h'.
\eal 
Hence, subtracting the same quantity with $\h$ and $\h'$ interchanged we obtain
\bal
&\hskip-2.cm-\frac12F^{\n\k}\lbar\h\g^{\r}{}_{\n\k}\g^5 \cd_\r\h'+\frac12F^{\n\k}\lbar\h'\g^{\r}{}_{\n\k}\g^5 \cd_\r\h\NO\\
=&\;-\frac12F^{\n\k}\lbar\h\g^{\r}{}_{\n\k}\g^5 \cd_\r\h'+\frac12F^{\n\k}\de_\r\big(\lbar\h'\g^{\r}{}_{\n\k}\g^5\h\big)+\frac12F^{\n\k}\lbar\h\g^5\g_{\k\n}{}^\r \cd_\r\h'\NO\\
=&\;\frac12F^{\n\k}\de_\r\big(\lbar\h'\g^{\r}{}_{\n\k}\g^5\h\big),
\eal
which again vanishes up to a total derivative term due to the Bianchi identity $\e^{\m\n\r\s}\pa_\r F_{\m\n}=0$ .

In order to evaluate the term proportional to the Schouten tensor $P_{\m\n}$
we note that
\bal
&\hskip-2.cm\lbar\h\g^{\m\n}\g_\l \cd_\k\h'-\lbar\h'\g^{\m\n}\g_\l \cd_\k\h\NO\\
=&\;\lbar\h\{\g^{\m\n},\g_{\l}\}\cd_\k\h'-\de_\k(\lbar\h'\g^{\m\n}\g_\l\h)\NO\\
=&\;2\lbar\h\g^{\m\n}{}_{\l}\cd_\k\h'-\de_\k\big(\lbar\h'(\g^{\m\n}{}_\l+\g^{\m}g^{\n}_\l-\g^{\n}g^\m_\l)\h\big).
\eal
Hence,
\bal
&P_{\m\n}\big(\lbar\h g^{\m[\n}\g^{\k\l]}\g_\l \cd_\k\h'-\lbar\h' g^{\m[\n}\g^{\k\l]}\g_\l \cd_\k\h\big)\NO\\
&=\frac{1}{3}P_{\m\n}\big(g^{\m\n}\lbar\h\g^{\k\l}\g_\l \cd_\k\h'+g^{\m\l}\lbar\h\g^{\n\k}\g_\l \cd_\k\h'+g^{\m\k}\lbar\h\g^{\l\n}\g_\l \cd_\k\h'-\h'\leftrightarrow\h\big)\NO\\
&=-\frac{1}{3}P_{\m\n}\de_\k\big(g^{\m\n}\lbar\h'(\g^{\k}g^\l_\l-\g^\l g^\k_\l)\h+g^{\m\l}\lbar\h'(\g^{\n}g^\k_\l-\g^\k g^\n_\l)\h+g^{\m\k}\lbar\h'(\g^{\l}g^\n_\l-\g^\n g^\l_\l)\h\big)\NO\\
&=-\frac{1}{3}P_{\m\n}\de_\k\big(3g^{\m\n}\lbar\h'\g^\k\h+\lbar\h'(\g^{\n}g^{\m\k}-\g^\k g^{\m\n})\h-3g^{\m\k}\lbar\h'\g^\n\h\big)\NO\\
&=-\frac{2}{3}P_{\m\n}\de_\k\big(g^{\m\n}\lbar\h'\g^\k\h-g^{\m\k}\lbar\h'\g^\n\h\big)\NO\\
&=-\frac{1}{6}R\de_\m(\lbar\h'\g^\m\h)+\frac13R_{\m\n}\de^\m(\lbar\h'\g^\n\h)=\frac13\de^\m\Big[\Big(R_{\m\n}-\frac12 R g_{\m\n}\Big)(\lbar\h'\g^\n\h)\Big],
\eal
where in the last equality we have used the Bianchi identity 
\be\label{Bianchi-Einstein}
\de^\m\Big(R_{\m\n}-\frac12 R g_{\m\n}\Big)=0.
\ee
It follows that the term proportional to the Schouten tensor $P_{\m\n}$ in \eqref{S^2} also vanishes.

Finally, for the last term in \eqref{S^2} we have 
\be
R^{\m\n\r\s}\lbar\h\g_{\m\n}\g_\s \cd_\r\h'=R^{\m\n\r\s}\lbar\h(\cancelto{0}{\g_{\m\n\s}}+\g_\m g_{\n\s}-\g_\n g_{\m\s})\cd_\r\h'=2R^{\m\r}\lbar\h\g_\m \cd_\r\h',
\ee
with
\be
R^{\m\n}\lbar\h\g_\m \cd_\n\h'-R^{\m\n}\lbar\h'\g_\m \cd_\n\h=R^{\m\n}\de_\n(\lbar\h\g_\m \h').
\ee
Hence,
\bal
&\hskip-2.cmR^{\m\n\r\s}\lbar\h\g_{\m\n}\g_\s \cd_\r\h'-R^{\m\n\r\s}\lbar\h'\g_{\m\n}\g_\s \cd_\r\h-\frac12 Rg_{\m\n}\big(\lbar\h g^{\m[\n}\g^{\k\l]}\g_\l \cd_\k\h'-\lbar\h' g^{\m[\n}\g^{\k\l]}\g_\l \cd_\k\h\big)\NO\\
=&\;2R^{\m\n}\de_\n(\lbar\h\g_\m \h')+R\de_\m(\lbar\h'\g^\m\h)=2\de^\m\Big[\Big(R_{\m\n}-\frac12 R g_{\m\n}\Big)(\lbar\h\g^\n\h')\Big],
\eal
where we have again used the Bianchi identity \eqref{Bianchi-Einstein} in the last equality. 

To summarize, all terms in \eqref{S^2} vanish upon subtracting the same quantities with $\h$ and $\h'$ interchanged, leading to the commutator
\be
[\d_\h,\d_{\h'}]\mathscr{W}=0,
\ee
in agreement with the Wess-Zumino consistency conditions.

\begin{flushleft}
\begin{tabular}{l}
\hline\hline
\raisebox{0.1cm}{\mbox{$[\d_{\ve}, \d_{\h}]\mathscr{W}=\d_\s\mathscr{W}+\d_\th\mathscr{W}$, with $\s=\frac12\lbar\ve\h$ and $\th=-\frac{3i}{4}\lbar\ve\g^5\h$:\rule{0cm}{.5cm}\hspace{6.51cm}}}\\
\hline\hline
\end{tabular}
\end{flushleft}

The final commutator we need to consider is the one between $\cq$- and $\cs$-supersymmetry transformations. We start with the contribution of the $\ca_Q$ anomaly to the commutator, namely 
\bal\label{dSdQ}
\d_{\h} \d_{\ve}\mathscr{W}=&\;-\d_{\h}\int d^4x\;e\;\lbar\ve\ca_{Q}\NO\\
=&\;\frac{(5a-3c)i}{18\p^2}\int d^4 x\;e\;\e^{\m\n\r\s}F_{\m\n}A_\r\lbar\ve\g^5\d_{\h}\f_\s-\frac{(a-c)}{12\p^2}\int d^4x\;e\;\e^{\m\n\r\s}\nabla_\k\big(A_\r R^{\k\l}{}_{\m\n} \big)\lbar\ve\g_{(\l}\d_\h\j_{\s)}\NO\\
&+\frac{(a-c)}{48\p^2}\int d^4x\;e\;\e^{\m\n\r\s}F_{\r\s} R^{\k\l}{}_{\m\n} \lbar\ve\g_\k\d_\h\j_\l\NO\\
=&\;\frac{(5a-3c)i}{18\p^2}\int d^4 xe\;\e^{\m\n\r\s}F_{\m\n}A_\r\lbar\ve\g^5\cd_\s\h+\frac{(a-c)}{12\p^2}\int d^4x\;e\;\cancelto{0}{\e^{\m\n\r\s}\nabla_\k\big(A_\r R^{\k}{}_{\s\m\n} \big)}\lbar\ve\h\NO\\
&-\frac{(a-c)}{48\p^2}\int d^4x\;e\;\e^{\m\n\r\s}F_{\r\s} R^{\k\l}{}_{\m\n} \lbar\ve\g_{\k\l}\h\NO\\
=&\;\frac{(5a-3c)i}{18\p^2}\int d^4 x
\;e\;\e^{\m\n\r\s}F_{\m\n}A_\r\lbar\ve\g^5\cd_\s\h-\frac{(a-c)}{48\p^2}\int d^4x\;e\;\e^{\m\n\r\s}F_{\r\s} R^{\k\l}{}_{\m\n} \lbar\ve\g_{\k\l}\h.
\eal
The last term in this expression can be simplified further using the last identity in \eqref{d=4-gamma-ids} and the product of two Levi-Civita tensors  
\bal
&\hskip-1.5cm-\e^{\m\n\r\s}\e^{\k\l\t\f}\NO\\
=&\;\blue{g^{\m\k}g^{\n\l}g^{\r\t}g^{\s\f}}+\big(\blue{g^{\m\l}g^{\n\k}g^{\r\f}g^{\s\t}}+\red{g^{\m\t}g^{\n\f}g^{\r\k}g^{\s\l}+g^{\m\f}g^{\n\t}g^{\r\l}g^{\s\k}}\big)-\big(\blue{g^{\m\l}g^{\n\k}g^{\r\t}g^{\s\f}}\NO\\
&+\textcolor{orange}{g^{\m\t}g^{\n\l}g^{\r\k}g^{\s\f}}+\textcolor{olive}{g^{\m\f}g^{\n\l}g^{\r\t}g^{\s\k}}+\textcolor{purple}{g^{\m\k}g^{\n\t}g^{\r\l}g^{\s\f}}+\textcolor{teal}{g^{\m\k}g^{\n\f}g^{\r\t}g^{\s\l}}+\blue{g^{\m\k}g^{\n\l}g^{\r\f}g^{\s\t}}\big)\NO\\
&+\big(\textcolor{orange}{g^{\m\t}g^{\n\k}g^{\r\l}g^{\s\f}}+\textcolor{purple}{g^{\m\l}g^{\n\t}g^{\r\k}g^{\s\f}}+\textcolor{olive}{g^{\m\f}g^{\n\k}g^{\r\t}g^{\s\l}}+\textcolor{teal}{g^{\m\l}g^{\n\f}g^{\r\t}g^{\s\k}}+\textcolor{orange}{g^{\m\f}g^{\n\l}g^{\r\k}g^{\s\t}}\NO\\
&+\textcolor{olive}{g^{\m\t}g^{\n\l}g^{\r\f}g^{\s\k}}+\textcolor{purple}{g^{\m\k}g^{\n\f}g^{\r\l}g^{\s\t}}+\textcolor{teal}{g^{\m\k}g^{\n\t}g^{\r\f}g^{\s\l}}\big)-\big(\textcolor{orange}{g^{\m\f}g^{\n\k}g^{\r\l}g^{\s\t}}+\textcolor{olive}{g^{\m\t}g^{\n\k}g^{\r\f}g^{\s\l}}\NO\\
&+\red{g^{\m\f}g^{\n\t}g^{\r\k}g^{\s\l}}+\textcolor{purple}{g^{\m\l}g^{\n\f}g^{\r\k}g^{\s\t}}+\red{g^{\m\t}g^{\n\f}g^{\r\l}g^{\s\k}}+\textcolor{teal}{g^{\m\l}g^{\n\t}g^{\r\f}g^{\s\k}}\big),
\eal
where we have grouped terms according to the conjugacy classes of the symmetric group $S_4$ and terms of a given color give the same result when contracted with $F_{\m\n}R_{\r\s\k\l}\lbar\h\g_{\t\f}\g^5\ve$. In particular,
\bal\label{ee-term}
&\hskip-2.cm-\e^{\m\n\r\s}\e^{\k\l\t\f}F_{\m\n}R_{\r\s\k\l}\lbar\h\g_{\t\f}\g^5\ve\NO\\
=&\;\blue{4F^{\m\n}R_{\m\n\r\s}\lbar\h\g^{\r\s}\g^5\ve}-\textcolor{orange}{4F_{\m}{}^\n R_{\n\r}\lbar\h\g^{\m\r}\g^5\ve}+\textcolor{olive}{4F_{\m}{}^\n R_{\r\n}\lbar\h\g^{\r\m}\g^5\ve}\NO\\
&+\textcolor{purple}{4F^{\n}{}_\m R_{\n\r}\lbar\h\g^{\m\r}\g^5\ve}-\textcolor{teal}{4F^\n{}_{\m}R_{\r\n}\lbar\h\g^{\r\m}\g^5\ve}+\red{4F_{\m\n}R\lbar\h\g^{\m\n}\g^5\ve}\NO\\
=&\;4\big(F^{\m\n}R_{\m\n\r\s}\lbar\h\g^{\r\s}\g^5\ve-4F_{\m}{}^\n R_{\n\r}\lbar\h\g^{\m\r}\g^5\ve+F_{\m\n}R\lbar\h\g^{\m\n}\g^5\ve\big),
\eal
and so
\bal\label{SQ-1}
\e^{\m\n\r\s}F_{\r\s} R^{\k\l}{}_{\m\n}\lbar\h\g_{\k\l}\ve=&\;\frac{i}{2}\e^{\m\n\r\s}\e^{\k\l\t\f}F_{\m\n}R_{\r\s\k\l}\lbar\h\g_{\t\f}\g^5\ve\NO\\
=&\;-2i\big(F^{\m\n}R_{\m\n\r\s}\lbar\h\g^{\r\s}\g^5\ve-4F_{\m}{}^\n R_{\n\r}\lbar\h\g^{\m\r}\g^5\ve+F_{\m\n}R\lbar\h\g^{\m\n}\g^5\ve\big).
\eal

The contribution of the $\ca_S$ anomaly to the commutator is
\bal\label{dQdS}
\d_{\ve} \d_{\h}\mathscr{W}=&\;\d_{\ve}\int d^4x\;e\;\lbar\h\ca_{S}\NO\\
=&\;-\frac{(5a-3c)}{18\p^2}i\int d^4 x\;e\;\e^{\m\n\r\s}F_{\m\n}A_\r\lbar\h\g^5\d_{\ve}\j_\s+\frac{(5a-3c)}{12\p^2}\int d^4x\;e\;\e^{\m\n\r\s}F_{\m\n}\lbar\h \cd_\r\d_{\ve}\j_{\s}\NO\\
&+\frac{ic}{6\p^2} \int d^4x\;e\;F^{\m\n}\lbar\h\big(\g_{\m}{}^{[\s}\d_{\n}{}^{\r]}-\d_{\m}{}^{[\s}\d_{\n}{}^{\r]}\big)\g^5\cd_\r\d_{\ve}\j_\s+\frac{3(2a-c)}{4\p^2}\int d^4x\;e\;P_{\m\n}g^{\m[\n}\lbar\h\g^{\r\s]}\cd_\r\d_{\ve}\j_\s\NO\\
&+\frac{(a-c)}{8\p^2}\int d^4x\;e\;\lbar\h\Big(R^{\m\n\r\s}\g_{\m\n}-\frac12Rg_{\m\n}g^{\m[\n}\g^{\r\s]}\Big)\cd_\r\d_{\ve}\j_\s\NO\\
%%%%%%%%%%%%%%%%%%%%%%%%%%%%%%%%%%%%%%%%%%%%%%%%%%%%%%%%%%%%%%%%%%%%%%%%
=&\;-\frac{(5a-3c)}{18\p^2}i\int d^4 x\;e\;\e^{\m\n\r\s}F_{\m\n}A_\r\lbar\h\g^5\cd_\s\ve+\frac{(5a-3c)}{12\p^2}\int d^4x\;e\;\e^{\m\n\r\s}F_{\m\n}\lbar\h \cd_\r \cd_\s\ve\NO\\
&+\frac{ic}{6\p^2} \int d^4x\;e\;F^{\m\n}\lbar\h\big(\g_{\m}{}^{[\s}\d_{\n}{}^{\r]}-\d_{\m}{}^{[\s}\d_{\n}{}^{\r]}\big)\g^5\cd_\r \cd_\s\ve+\frac{3(2a-c)}{4\p^2}\int d^4x\;e\;P_{\m\n}g^{\m[\n}\lbar\h\g^{\r\s]}\cd_\r \cd_\s\ve\NO\\
&+\frac{(a-c)}{8\p^2}\int d^4x\;e\;\lbar\h\Big(R^{\m\n\r\s}\g_{\m\n}-\frac12Rg_{\m\n}g^{\m[\n}\g^{\r\s]}\Big)\cd_\r \cd_\s\ve.
\eal
Using the identity,
\be
2\cd_{[\m} \cd_{\n]}\ve=\Big(\frac14 R_{\m\n\r\s}\g^{\r\s}+i\g^5F_{\m\n}\Big)\ve,
\ee
we now evaluate the four terms involving two covariant derivatives on the spinor parameter $\ve$. The easiest term to evaluate is
\bal\label{QS-1}
&\hskip-1.5cm iF^{\m\n}\lbar\h\big(\g_{\m}{}^{[\s}\d_{\n}{}^{\r]}-\d_{\m}{}^{[\s}\d_{\n}{}^{\r]}\big)\g^5\cd_\r \cd_\s\ve\NO\\
=&\;\frac{i}{2}F^{\m\n}\lbar\h\big(\g_{\m}{}^{\s}\d_{\n}{}^{\r}-\d_{\m}{}^{\s}\d_{\n}{}^{\r}\big)\g^5\Big(\frac14 R_{\r\s\k\l}\g^{\k\l}+i\g^5F_{\r\s}\Big)\ve\NO\\
=&\;-\frac{1}{2}F_{\m\n}F^{\m\n}\lbar\h\ve+\frac{i}{8}F_{\m}{}^\n R_{\n\s\k\l}\lbar\h\g^{\m\s}\g^{\k\l}\g^5\ve+\frac{i}{8}F^{\m\n}R_{\m\n\k\l}\lbar\h\g^{\k\l}\g^5\ve\NO\\
=&\;-\frac{1}{2}F_{\m\n}F^{\m\n}\lbar\h\ve+\frac{i}{8}F^{\m\n}R_{\m\n\k\l}\lbar\h\g^{\k\l}\g^5\ve\NO\\
&+\frac{i}{8}F_{\m}{}^\n R_{\n\s\k\l}\lbar\h\big(-i\cancelto{0}{\e^{\m\s\k\l}\g^5}+2\g^{\m\l}g^{\s\k}-2\g^{\s\l}g^{\m\k}+\cancelto{0}{g^{\m\l}g^{\s\k}}-\cancelto{0}{g^{\s\l}g^{\m\k}}\big)\g^5\ve\NO\\
=&\;-\frac{1}{2}F_{\m\n}F^{\m\n}\lbar\h\ve+\frac{i}{8}F^{\m\n}(R_{\m\n\r\s}+2 R_{\m\r\n\s}-2g_{\m\r}R_{\n\s})\lbar\h\g^{\r\s}\g^5\ve\NO\\
=&\;-\frac{1}{2}F_{\m\n}F^{\m\n}\lbar\h\ve+\frac{i}{4}F^{\m\n}(R_{\m\n\r\s}-g_{\m\r}R_{\n\s})\lbar\h\g^{\r\s}\g^5\ve.
\eal

We next consider the term
\bal\label{QS-2}
&\hskip-1.5cm \e^{\m\n\r\s}F_{\m\n}\lbar\h \cd_\r \cd_\s\ve\NO\\
%=&\;\frac{i}{2}F^{\m\n}\lbar\h\big(-i\e_{\m\n}{}^{\r\s}\g^5\big)\g^5\Big(\frac14 R_{\r\s\k\l}\g^{\k\l}+i\g^5F_{\r\s}\Big)\ve\NO\\
=&\;\frac{i}{2}\e^{\m\n\r\s}F_{\m\n}F_{\r\s}\lbar\h\g^5\ve+\frac{1}{8}\e^{\m\n\r\s}F_{\m\n}R_{\r\s\k\l}\lbar\h\g^{\k\l}\ve\NO\\
=&\;\frac{i}{2}\e^{\m\n\r\s}F_{\m\n}F_{\r\s}\lbar\h\g^5\ve-\frac{i}{4}\big(F^{\m\n}R_{\m\n\r\s}\lbar\h\g^{\r\s}\g^5\ve-4F_{\m}{}^\n R_{\n\r}\lbar\h\g^{\m\r}\g^5\ve+F_{\m\n}R\lbar\h\g^{\m\n}\g^5\ve\big),
\eal
where in the last line we have utilized the identity \eqref{SQ-1}.

Next we evaluate the term
\bal\label{QS-3}
&\hskip-1.5cm P_{\m\n} g^{\m[\n}\lbar\h\g^{\k\l]}\cd_\k \cd_\l\ve\NO\\
=&\;\frac{1}{6}P_{\m\n}\lbar\h \big(g^{\m\n}\g^{\k\l}+g^{\m\l}\g^{\n\k}+g^{\m\k}\g^{\l\n}\big)\Big(\frac14 R_{\k\l\r\s}\g^{\r\s}+i\g^5F_{\k\l}\Big)\ve\NO\\
=&\;\frac{1}{6}P_{\m\n}\lbar\h (g^{\m\n}\g^{\k\l}-2g^{\m\k}\g^{\n\l})\Big(\frac14 R_{\k\l\r\s}\g^{\r\s}+i\g^5F_{\k\l}\Big)\ve\NO\\
=&\;\frac{1}{36}R\Big(\frac14 R_{\k\l\r\s}\lbar\h\g^{\k\l}\g^{\r\s}\ve+i\lbar\h\g^{\k\l}\g^5\ve F_{\k\l}\Big)-\frac13P_{\m\n}\Big(\frac14 R^\m{}_{\l\r\s}\lbar\h \g^{\n\l}\g^{\r\s}\ve+i\lbar\h \g^{\n\l}\g^5\ve F^\m{}_{\l}\Big)\NO\\
=&\;-\frac{1}{6}\Big(R_{\m\n}-\frac{1}{3}Rg_{\m\n}\Big)\times\NO\\
&\Big(iF^\m{}_{\l}\lbar\h \g^{\n\l}\g^5\ve+\frac{1}{4} R^\m{}_{\l\r\s}\lbar\h \big(-i\cancelto{0}{\e^{\n\l\r\s}}\g^5+2\cancelto{0}{\g^{\n\s}g^{\l\r}}-2\cancelto{0}{\g^{\l\s}g^{\n\r}}+g^{\n\s}g^{\l\r}-g^{\l\s}g^{\n\r}\big)\ve\Big)\NO\\
=&\;-\frac{1}{6}\Big(R_{\m\n}-\frac{1}{3}Rg_{\m\n}\Big)iF^\m{}_{\l}\lbar\h \g^{\n\l}\g^5\ve+\frac{1}{12}\Big(R_{\m\n}R^{\m\n}-\frac{1}{3}R^2\Big) \lbar\h\ve.
\eal

Finally, the last term gives
\bal\label{QS-4}
&\hskip-1.cm\Big(R^{\m\n\r\s}\g_{\m\n}-\frac12Rg_{\m\n}g^{\m[\n}\g^{\r\s]}\Big)\cd_\r \cd_\s\ve\NO\\
=&\;\frac12\Big(R^{\m\n\r\s}\g_{\m\n}-\frac12Rg_{\m\n}g^{\m[\n}\g^{\r\s]}\Big)\Big(\frac14 R_{\r\s\k\l}\g^{\k\l}+i\g^5F_{\r\s}\Big)\ve\NO\\
=&\;\frac{1}{8}R^{\m\n\r\s}R_{\k\l\r\s}\big(-i\e_{\m\n}{}^{\k\l}\g^5+4\cancelto{0}{\g_\m{}^\l g_\n^\k}+2g_\m^\l g_\n^\k\big)\ve\NO\\
&+\frac{i}{2}R^{\m\n\r\s}F_{\r\s}\g_{\m\n}\g^5\ve-\frac{1}{12}\big(2iRF_{\m\n}\lbar\h \g^{\m\n}\g^5\ve-R^2 \lbar\h\ve\big)\NO\\
=&\;-\frac{i}{4}\cp\g^5\ve-\frac{1}{4}R_{\m\n\r\s}R^{\m\n\r\s}\ve+\frac{i}{2}R^{\m\n\r\s}F_{\r\s}\g_{\m\n}\g^5\ve-\frac{1}{12}\big(2iRF_{\m\n}\lbar\h \g^{\m\n}\g^5\ve-R^2 \lbar\h\ve\big),
\eal
where $\cp=\frac12\e^{\m\n\r\s}R_{\m\n\k\l}R_{\r\s}{}^{\k\l}$ is the Pontryagin density.

Substituting \eqref{SQ-1} in \eqref{dSdQ}, and \eqref{QS-1}, \eqref{QS-2}, \eqref{QS-3} and \eqref{QS-4} in \eqref{dQdS} we finally obtain
\bal
[\d_{\ve}, \d_{\h}]\mathscr{W}=&\;\frac{(5a-3c)i}{72\p^2}\int d^4x\;e\;\e^{\m\n\r\s}F_{\m\n}F_{\r\s}\lbar\h\g^5\ve-\frac{c}{12\p^2} \int d^4x\;e\;F_{\m\n}F^{\m\n}\lbar\h\ve\NO\\
&+\frac{(2a-c)}{16\p^2}\int d^4x\;e\;\Big(R_{\m\n}R^{\m\n}-\frac{1}{3}R^2\Big) \lbar\h\ve\NO\\
&-\frac{(a-c)}{32\p^2}\int d^4x\;e\;\Big(i\cp\lbar\h\g^5\ve+R_{\m\n\r\s}R^{\m\n\r\s}\lbar\h\ve-\frac{1}{3}R^2\lbar\h\ve\Big)\NO\\
%%%%%%%%%%%%%%%%%%%%%%%%%%%%%%%%%%%%%%%%%%%%%%%%%%%%%%%%%%%%%%%%%%%%%%%%%%%%%%%%
=&\;\int d^4 x\;e\;(-\th\ca_R+\s \ca_W),
\eal
with
\be
\th=-\frac34i\lbar\ve\g^5\h,\qquad \s=\frac12\lbar\ve\h,
\ee
as required by the Wess-Zumino consistency conditions.

%\nocite{*}

%\bibliographystyle{plain}
%\bibliographystyle{hieeetr}
\bibliographystyle{jhepcap}
\bibliography{CSanomalies}

\providecommand{\href}[2]{#2}\begingroup\raggedright\begin{thebibliography}{10}

\bibitem{Witten:1982im}
E.~Witten, {\it {Supersymmetry and Morse theory}},  {\em J. Diff. Geom.} {\bf
  17} (1982), no.~4 661--692.

\bibitem{Witten:1988ze}
E.~Witten, {\it {Topological Quantum Field Theory}},  {\em Commun. Math. Phys.}
  {\bf 117} (1988) 353.

\bibitem{Nekrasov:2002qd}
N.~A. Nekrasov, {\it {Seiberg-Witten prepotential from instanton counting}},
  {\em Adv. Theor. Math. Phys.} {\bf 7} (2003), no.~5 831--864,
  [\href{http://xxx.lanl.gov/abs/hep-th/0206161}{{\tt hep-th/0206161}}].

\bibitem{Pestun:2007rz}
V.~Pestun, {\it {Localization of gauge theory on a four-sphere and
  supersymmetric Wilson loops}},  {\em Commun. Math. Phys.} {\bf 313} (2012)
  71--129, [\href{http://xxx.lanl.gov/abs/0712.2824}{{\tt 0712.2824}}].

\bibitem{Wess:1971yu}
J.~Wess and B.~Zumino, {\it {Consequences of anomalous Ward identities}},  {\em
  Phys. Lett.} {\bf 37B} (1971) 95--97.

\bibitem{Katsianis:2019hhg}
G.~Katsianis, I.~Papadimitriou, K.~Skenderis, and M.~Taylor, {\it {Anomalous
  Supersymmetry}},  \href{http://xxx.lanl.gov/abs/1902.06715}{{\tt
  1902.06715}}.

\bibitem{An:2019zok}
O.~S. An, J.~U. Kang, J.~C. Kim, and Y.~H. Ko, {\it {Quantum consistency in
  supersymmetric theories with $R$-symmetry in curved space}},
  \href{http://xxx.lanl.gov/abs/1902.04525}{{\tt 1902.04525}}.

\bibitem{Papadimitriou:2017kzw}
I.~Papadimitriou, {\it {Supercurrent anomalies in 4d SCFTs}},  {\em JHEP} {\bf
  07} (2017) 038, [\href{http://xxx.lanl.gov/abs/1703.04299}{{\tt
  1703.04299}}].

\bibitem{An:2017ihs}
O.~S. An, {\it {Anomaly-corrected supersymmetry algebra and supersymmetric
  holographic renormalization}},  {\em JHEP} {\bf 12} (2017) 107,
  [\href{http://xxx.lanl.gov/abs/1703.09607}{{\tt 1703.09607}}].

\bibitem{An:2018roi}
O.~S. An, Y.~H. Ko, and S.-H. Won, {\it {Super-Weyl Anomaly from Holography and
  Rigid Supersymmetry Algebra on Two-Sphere}},
  \href{http://xxx.lanl.gov/abs/1812.10209}{{\tt 1812.10209}}.

\bibitem{Adler:1969gk}
S.~L. Adler, {\it {Axial vector vertex in spinor electrodynamics}},  {\em Phys.
  Rev.} {\bf 177} (1969) 2426--2438.

\bibitem{Bell:1969ts}
J.~S. Bell and R.~Jackiw, {\it {A PCAC puzzle: $\pi^0 \to \gamma \gamma$ in the
  $\sigma$ model}},  {\em Nuovo Cim.} {\bf A60} (1969) 47--61.

\bibitem{Landsteiner:2016led}
K.~Landsteiner, {\it {Notes on Anomaly Induced Transport}},  {\em Acta Phys.
  Polon.} {\bf B47} (2016) 2617,
  [\href{http://xxx.lanl.gov/abs/1610.04413}{{\tt 1610.04413}}].

\bibitem{Gooth:2017mbd}
J.~Gooth {\em et.~al.}, {\it {Experimental signatures of the mixed
  axial-gravitational anomaly in the Weyl semimetal NbP}},  {\em Nature} {\bf
  547} (2017) 324--327, [\href{http://xxx.lanl.gov/abs/1703.10682}{{\tt
  1703.10682}}].

\bibitem{Pestun:2016zxk}
V.~Pestun {\em et.~al.}, {\it {Localization techniques in quantum field
  theories}},  {\em J. Phys.} {\bf A50} (2017), no.~44 440301,
  [\href{http://xxx.lanl.gov/abs/1608.02952}{{\tt 1608.02952}}].

\bibitem{Festuccia:2011ws}
G.~Festuccia and N.~Seiberg, {\it {Rigid Supersymmetric Theories in Curved
  Superspace}},  {\em JHEP} {\bf 06} (2011) 114,
  [\href{http://xxx.lanl.gov/abs/1105.0689}{{\tt 1105.0689}}].

\bibitem{Samtleben:2012gy}
H.~Samtleben and D.~Tsimpis, {\it {Rigid supersymmetric theories in 4d
  Riemannian space}},  {\em JHEP} {\bf 05} (2012) 132,
  [\href{http://xxx.lanl.gov/abs/1203.3420}{{\tt 1203.3420}}].

\bibitem{Klare:2012gn}
C.~Klare, A.~Tomasiello, and A.~Zaffaroni, {\it {Supersymmetry on Curved Spaces
  and Holography}},  {\em JHEP} {\bf 08} (2012) 061,
  [\href{http://xxx.lanl.gov/abs/1205.1062}{{\tt 1205.1062}}].

\bibitem{Dumitrescu:2012ha}
T.~T. Dumitrescu, G.~Festuccia, and N.~Seiberg, {\it {Exploring Curved
  Superspace}},  {\em JHEP} {\bf 08} (2012) 141,
  [\href{http://xxx.lanl.gov/abs/1205.1115}{{\tt 1205.1115}}].

\bibitem{Liu:2012bi}
J.~T. Liu, L.~A. Pando~Zayas, and D.~Reichmann, {\it {Rigid Supersymmetric
  Backgrounds of Minimal Off-Shell Supergravity}},  {\em JHEP} {\bf 10} (2012)
  034, [\href{http://xxx.lanl.gov/abs/1207.2785}{{\tt 1207.2785}}].

\bibitem{Dumitrescu:2012at}
T.~T. Dumitrescu and G.~Festuccia, {\it {Exploring Curved Superspace (II)}},
  {\em JHEP} {\bf 01} (2013) 072,
  [\href{http://xxx.lanl.gov/abs/1209.5408}{{\tt 1209.5408}}].

\bibitem{Kehagias:2012fh}
A.~Kehagias and J.~G. Russo, {\it {Global Supersymmetry on Curved Spaces in
  Various Dimensions}},  {\em Nucl. Phys.} {\bf B873} (2013) 116--136,
  [\href{http://xxx.lanl.gov/abs/1211.1367}{{\tt 1211.1367}}].

\bibitem{Closset:2012ru}
C.~Closset, T.~T. Dumitrescu, G.~Festuccia, and Z.~Komargodski, {\it
  {Supersymmetric Field Theories on Three-Manifolds}},  {\em JHEP} {\bf 05}
  (2013) 017, [\href{http://xxx.lanl.gov/abs/1212.3388}{{\tt 1212.3388}}].

\bibitem{Samtleben:2012ua}
H.~Samtleben, E.~Sezgin, and D.~Tsimpis, {\it {Rigid 6D supersymmetry and
  localization}},  {\em JHEP} {\bf 03} (2013) 137,
  [\href{http://xxx.lanl.gov/abs/1212.4706}{{\tt 1212.4706}}].

\bibitem{Cassani:2012ri}
D.~Cassani, C.~Klare, D.~Martelli, A.~Tomasiello, and A.~Zaffaroni, {\it
  {Supersymmetry in Lorentzian Curved Spaces and Holography}},  {\em Commun.
  Math. Phys.} {\bf 327} (2014) 577--602,
  [\href{http://xxx.lanl.gov/abs/1207.2181}{{\tt 1207.2181}}].

\bibitem{deMedeiros:2012sb}
P.~de~Medeiros, {\it {Rigid supersymmetry, conformal coupling and twistor
  spinors}},  {\em JHEP} {\bf 09} (2014) 032,
  [\href{http://xxx.lanl.gov/abs/1209.4043}{{\tt 1209.4043}}].

\bibitem{Hristov:2013spa}
K.~Hristov, A.~Tomasiello, and A.~Zaffaroni, {\it {Supersymmetry on
  Three-dimensional Lorentzian Curved Spaces and Black Hole Holography}},  {\em
  JHEP} {\bf 05} (2013) 057, [\href{http://xxx.lanl.gov/abs/1302.5228}{{\tt
  1302.5228}}].

\bibitem{Blau:2000xg}
M.~Blau, {\it {Killing spinors and SYM on curved spaces}},  {\em JHEP} {\bf 11}
  (2000) 023, [\href{http://xxx.lanl.gov/abs/hep-th/0005098}{{\tt
  hep-th/0005098}}].

\bibitem{Kuzenko:2012vd}
S.~M. Kuzenko, {\it {Symmetries of curved superspace}},  {\em JHEP} {\bf 03}
  (2013) 024, [\href{http://xxx.lanl.gov/abs/1212.6179}{{\tt 1212.6179}}].

\bibitem{Closset:2013vra}
C.~Closset, T.~T. Dumitrescu, G.~Festuccia, and Z.~Komargodski, {\it {The
  Geometry of Supersymmetric Partition Functions}},  {\em JHEP} {\bf 01} (2014)
  124, [\href{http://xxx.lanl.gov/abs/1309.5876}{{\tt 1309.5876}}].

\bibitem{Closset:2014uda}
C.~Closset, T.~T. Dumitrescu, G.~Festuccia, and Z.~Komargodski, {\it {From
  Rigid Supersymmetry to Twisted Holomorphic Theories}},  {\em Phys. Rev.} {\bf
  D90} (2014), no.~8 085006, [\href{http://xxx.lanl.gov/abs/1407.2598}{{\tt
  1407.2598}}].

\bibitem{Assel:2014paa}
B.~Assel, D.~Cassani, and D.~Martelli, {\it {Localization on Hopf surfaces}},
  {\em JHEP} {\bf 08} (2014) 123,
  [\href{http://xxx.lanl.gov/abs/1405.5144}{{\tt 1405.5144}}].

\bibitem{Genolini:2016ecx}
P.~Benetti~Genolini, D.~Cassani, D.~Martelli, and J.~Sparks, {\it {Holographic
  renormalization and supersymmetry}},  {\em JHEP} {\bf 02} (2017) 132,
  [\href{http://xxx.lanl.gov/abs/1612.06761}{{\tt 1612.06761}}].

\bibitem{Henningson:1998gx}
M.~Henningson and K.~Skenderis, {\it {The Holographic Weyl anomaly}},  {\em
  JHEP} {\bf 07} (1998) 023,
  [\href{http://xxx.lanl.gov/abs/hep-th/9806087}{{\tt hep-th/9806087}}].

\bibitem{deHaro:2000vlm}
S.~de~Haro, S.~N. Solodukhin, and K.~Skenderis, {\it {Holographic
  reconstruction of space-time and renormalization in the AdS / CFT
  correspondence}},  {\em Commun. Math. Phys.} {\bf 217} (2001) 595--622,
  [\href{http://xxx.lanl.gov/abs/hep-th/0002230}{{\tt hep-th/0002230}}].

\bibitem{McArthur:1983fk}
I.~N. McArthur, {\it {Super $b$(4) Coefficients in Supergravity}},  {\em Class.
  Quant. Grav.} {\bf 1} (1984) 245.

\bibitem{Bonora:1984pn}
L.~Bonora, P.~Pasti, and M.~Tonin, {\it {Cohomologies and Anomalies in
  Supersymmetric Theories}},  {\em Nucl. Phys.} {\bf B252} (1985) 458--480.

\bibitem{Buchbinder:1986im}
I.~L. Buchbinder and S.~M. Kuzenko, {\it {Matter Superfields in External
  Supergravity: Green Functions, Effective Action and Superconformal
  Anomalies}},  {\em Nucl. Phys.} {\bf B274} (1986) 653--684.

\bibitem{Brandt:1993vd}
F.~Brandt, {\it {Anomaly candidates and invariants of D = 4, N=1 supergravity
  theories}},  {\em Class. Quant. Grav.} {\bf 11} (1994) 849--864,
  [\href{http://xxx.lanl.gov/abs/hep-th/9306054}{{\tt hep-th/9306054}}].

\bibitem{Anselmi:1997am}
D.~Anselmi, D.~Z. Freedman, M.~T. Grisaru, and A.~A. Johansen, {\it
  {Nonperturbative formulas for central functions of supersymmetric gauge
  theories}},  {\em Nucl. Phys.} {\bf B526} (1998) 543--571,
  [\href{http://xxx.lanl.gov/abs/hep-th/9708042}{{\tt hep-th/9708042}}].

\bibitem{Piguet:1998bj}
O.~Piguet and S.~Wolf, {\it {The Supercurrent trace identities of the N=1, D =
  4 superYang-Mills theory in the Wess-Zumino gauge}},  {\em JHEP} {\bf 04}
  (1998) 001, [\href{http://xxx.lanl.gov/abs/hep-th/9802027}{{\tt
  hep-th/9802027}}].

\bibitem{Erdmenger:1998ew}
J.~Erdmenger and C.~Rupp, {\it {Geometrical superconformal anomalies}},
  \href{http://xxx.lanl.gov/abs/hep-th/9809090}{{\tt hep-th/9809090}}.

\bibitem{Erdmenger:1998xv}
J.~Erdmenger and C.~Rupp, {\it {Superconformal Ward identities for Green
  functions with multiple supercurrent insertions}},  {\em Annals Phys.} {\bf
  276} (1999) 152--187, [\href{http://xxx.lanl.gov/abs/hep-th/9811209}{{\tt
  hep-th/9811209}}].

\bibitem{Bonora:2013rta}
L.~Bonora and S.~Giaccari, {\it {Weyl transformations and trace anomalies in
  N=1, D=4 supergravities}},  {\em JHEP} {\bf 08} (2013) 116,
  [\href{http://xxx.lanl.gov/abs/1305.7116}{{\tt 1305.7116}}].

\bibitem{Butter:2013ura}
D.~Butter and S.~M. Kuzenko, {\it {Nonlocal action for the super-Weyl
  anomalies: A new representation}},  {\em JHEP} {\bf 09} (2013) 067,
  [\href{http://xxx.lanl.gov/abs/1307.1290}{{\tt 1307.1290}}].

\bibitem{Cassani:2013dba}
D.~Cassani and D.~Martelli, {\it {Supersymmetry on curved spaces and
  superconformal anomalies}},  {\em JHEP} {\bf 10} (2013) 025,
  [\href{http://xxx.lanl.gov/abs/1307.6567}{{\tt 1307.6567}}].

\bibitem{Auzzi:2015yia}
R.~Auzzi and B.~Keren-Zur, {\it {Superspace formulation of the local RG
  equation}},  {\em JHEP} {\bf 05} (2015) 150,
  [\href{http://xxx.lanl.gov/abs/1502.05962}{{\tt 1502.05962}}].

\bibitem{Wess:1992cp}
J.~Wess and J.~Bagger, {\em {Supersymmetry and supergravity}}.
\newblock Princeton University Press, Princeton, NJ, USA, 1992.

\bibitem{Piguet:1986ug}
O.~Piguet and K.~Sibold, {\em {Renormalized supersymmetry. The perturbation
  theory of N=1 supersymmetric theories in flat space-time}}, vol.~12.
\newblock 1986.

\bibitem{Itoyama:1985qi}
H.~Itoyama, V.~P. Nair, and H.-c. Ren, {\it {Supersymmetry Anomalies and Some
  Aspects of Renormalization}},  {\em Nucl. Phys.} {\bf B262} (1985) 317--330.

\bibitem{Guadagnini:1985ea}
E.~Guadagnini and M.~Mintchev, {\it {CHIRAL ANOMALIES AND SUPERSYMMETRY}},
  {\em Nucl. Phys.} {\bf B269} (1986) 543--556.

\bibitem{Piguet:1984aa}
O.~Piguet and K.~Sibold, {\it {The Anomaly in the Slavnov Identity for $N=1$
  Supersymmetric {Yang-Mills} Theories}},  {\em Nucl. Phys.} {\bf B247} (1984)
  484.

\bibitem{Kaku:1977pa}
M.~Kaku, P.~K. Townsend, and P.~van Nieuwenhuizen, {\it {Gauge Theory of the
  Conformal and Superconformal Group}},  {\em Phys. Lett.} {\bf 69B} (1977)
  304--308.

\bibitem{Kaku:1977rk}
M.~Kaku, P.~K. Townsend, and P.~van Nieuwenhuizen, {\it {Superconformal Unified
  Field Theory}},  {\em Phys. Rev. Lett.} {\bf 39} (1977) 1109.

\bibitem{Kaku:1978nz}
M.~Kaku, P.~K. Townsend, and P.~van Nieuwenhuizen, {\it {Properties of
  Conformal Supergravity}},  {\em Phys. Rev.} {\bf D17} (1978) 3179.
  [,853(1978)].

\bibitem{Townsend:1979ki}
P.~K. Townsend and P.~van Nieuwenhuizen, {\it {Simplifications of Conformal
  Supergravity}},  {\em Phys. Rev.} {\bf D19} (1979) 3166.

\bibitem{VanNieuwenhuizen:1981ae}
P.~Van~Nieuwenhuizen, {\it {Supergravity}},  {\em Phys. Rept.} {\bf 68} (1981)
  189--398.

\bibitem{deWit:1981vgr}
B.~de~Wit, {\it {CONFORMAL INVARIANCE IN EXTENDED SUPERGRAVITY}},  in {\em
  {First School on Supergravity Trieste, Italy, April 22-May 6, 1981}},
  p.~0267, 1981.

\bibitem{deWit:1983qkc}
B.~de~Wit, {\it {MULTIPLET CALCULUS}},  in {\em {September School on
  Supergravity and Supersymmetry Trieste, Italy, September 6-18, 1982}}, 1983.

\bibitem{Fradkin:1985am}
E.~S. Fradkin and A.~A. Tseytlin, {\it {CONFORMAL SUPERGRAVITY}},  {\em Phys.
  Rept.} {\bf 119} (1985) 233--362.

\bibitem{Freedman:2012zz}
D.~Z. Freedman and A.~Van~Proeyen, {\em {Supergravity}}.
\newblock Cambridge Univ. Press, Cambridge, UK, 2012.

\bibitem{Kaku:1978ea}
M.~Kaku and P.~K. Townsend, {\it {POINCARE SUPERGRAVITY AS BROKEN
  SUPERCONFORMAL GRAVITY}},  {\em Phys. Lett.} {\bf 76B} (1978) 54--58.

\bibitem{Stelle:1978yr}
K.~S. Stelle and P.~C. West, {\it {Tensor Calculus for the Vector Multiplet
  Coupled to Supergravity}},  {\em Phys. Lett.} {\bf 77B} (1978) 376.

\bibitem{Balasubramanian:2000pq}
V.~Balasubramanian, E.~G. Gimon, D.~Minic, and J.~Rahmfeld, {\it
  {Four-dimensional conformal supergravity from AdS space}},  {\em Phys. Rev.}
  {\bf D63} (2001) 104009, [\href{http://xxx.lanl.gov/abs/hep-th/0007211}{{\tt
  hep-th/0007211}}].

\bibitem{deWit:2018dix}
B.~de~Wit, S.~Murthy, and V.~Reys, {\it {BRST quantization and equivariant
  cohomology: localization with asymptotic boundaries}},  {\em JHEP} {\bf 09}
  (2018) 084, [\href{http://xxx.lanl.gov/abs/1806.03690}{{\tt 1806.03690}}].

\bibitem{Bardeen:1984pm}
W.~A. Bardeen and B.~Zumino, {\it {Consistent and Covariant Anomalies in Gauge
  and Gravitational Theories}},  {\em Nucl. Phys.} {\bf B244} (1984) 421--453.

\bibitem{Jensen:2012kj}
K.~Jensen, R.~Loganayagam, and A.~Yarom, {\it {Thermodynamics, gravitational
  anomalies and cones}},  {\em JHEP} {\bf 02} (2013) 088,
  [\href{http://xxx.lanl.gov/abs/1207.5824}{{\tt 1207.5824}}].

\bibitem{Nakayama:2012gu}
Y.~Nakayama, {\it {CP-violating CFT and trace anomaly}},  {\em Nucl. Phys.}
  {\bf B859} (2012) 288--298, [\href{http://xxx.lanl.gov/abs/1201.3428}{{\tt
  1201.3428}}].

\bibitem{Papadimitriou:2016yit}
I.~Papadimitriou, {\it {Lectures on Holographic Renormalization}},  {\em
  Springer Proc. Phys.} {\bf 176} (2016) 131--181.

\bibitem{Imbimbo:2014pla}
C.~Imbimbo and D.~Rosa, {\it {Topological anomalies for Seifert 3-manifolds}},
  {\em JHEP} {\bf 07} (2015) 068,
  [\href{http://xxx.lanl.gov/abs/1411.6635}{{\tt 1411.6635}}].

\bibitem{Kinney:2005ej}
J.~Kinney, J.~M. Maldacena, S.~Minwalla, and S.~Raju, {\it {An Index for 4
  dimensional super conformal theories}},  {\em Commun. Math. Phys.} {\bf 275}
  (2007) 209--254, [\href{http://xxx.lanl.gov/abs/hep-th/0510251}{{\tt
  hep-th/0510251}}].

\bibitem{Cabo-Bizet:2018ehj}
A.~Cabo-Bizet, D.~Cassani, D.~Martelli, and S.~Murthy, {\it {Microscopic origin
  of the Bekenstein-Hawking entropy of supersymmetric AdS$_{\bf 5}$ black
  holes}},  \href{http://xxx.lanl.gov/abs/1810.11442}{{\tt 1810.11442}}.

\bibitem{Choi:2018hmj}
S.~Choi, J.~Kim, S.~Kim, and J.~Nahmgoong, {\it {Large AdS black holes from
  QFT}},  \href{http://xxx.lanl.gov/abs/1810.12067}{{\tt 1810.12067}}.

\bibitem{Benini:2018ywd}
F.~Benini and P.~Milan, {\it {Black holes in 4d $\mathcal{N}=4$
  Super-Yang-Mills}},  \href{http://xxx.lanl.gov/abs/1812.09613}{{\tt
  1812.09613}}.

\bibitem{Honda:2019cio}
M.~Honda, {\it {Quantum Black Hole Entropy from 4d Supersymmetric Cardy
  formula}},  \href{http://xxx.lanl.gov/abs/1901.08091}{{\tt 1901.08091}}.

\bibitem{Bonora:1985cq}
L.~Bonora, P.~Pasti, and M.~Bregola, {\it {WEYL COCYCLES}},  {\em Class. Quant.
  Grav.} {\bf 3} (1986) 635.

\end{thebibliography}\endgroup

\end{document}